\documentclass[twocolumn,preprintnumbers,amsmath,amssymb,aps,prb]{revtex4}
\usepackage{graphicx}
\begin{document}

\title{
Individual Vortex Manipulation and Stick-Slip Motion in Periodic Pinning Arrays
}
\author{
Xiaoyu Ma$^{1,2}$, C. J. O. Reichhardt$^{1}$ and C. Reichhardt$^{1}$
}
\affiliation{
$^1$Theoretical Division,
Los Alamos National Laboratory, Los Alamos, New Mexico 87545 USA\\
$^2$ Department of Physics
University of Notre Dame, Notre Dame, Indiana 46556 USA
}

\date{\today}
\begin{abstract}
We numerically examine the manipulation of vortices interacting
with a moving trap representing
a magnetic force tip translating
across a superconducting sample containing a periodic array of pinning sites.
As a function of the tip velocity
and coupling strength, we find
five
distinct dynamic phases,
including a decoupled regime where the vortices are dragged a short distance within
a pinning site,
an intermediate coupling regime where
vortices in neighboring pinning sites exchange places,
an intermediate trapping regime where individual vortices are dragged longer
distances and
exchange modes of vortices occur
in the surrounding pins,
an intermittent multiple trapping regime where
the trap switches between capturing
one or two vortices,
and a strong couping regime in which the trap permanently
captures and drags two vortices.
In some regimes we
observe the counterintuitive behavior that
slow moving traps couple less strongly to vortices than faster moving traps; however,
the fastest moving traps are generally decoupled.
The different phases can
be characterized by the distances the
vortices are displaced and the force fluctuations
exerted on the trap.
We find
different types of stick-slip motion
depending on whether vortices are
moving into and out of pinning sites,
undergoing exchange,
or
performing correlated motion induced
by vortices outside of the trap.
Our results are
general to the manipulation of other types of particle-based
systems interacting with  periodic trap arrays,
such as colloidal particles or certain types of
frictional systems.
\end{abstract}
\maketitle

\section{Introduction}
Vortices in type-II superconductors interacting with ordered or disordered substrates
represent an outstanding example of a  condensed matter system
with competing interactions,
since the vortex-vortex repulsion favors a hexagonal lattice while the substrate
ordering can favor different lattice symmetries,
leading to commensurate-incommensurate
transitions \cite{1,2,3,4,5}, depinning phenomena in the presence of an external drive,
\cite{6,7,8,9,10}, and order-disorder transitions \cite{11,12,13}.
In addition to these basic science issues, vortex motion and pinning
are relevant to a variety of
applications such as critical current optimization \cite{11},
while there are a number of proposals for using
individual vortex manipulation to test aspects
of statistical physics \cite{14,15}
or to create new types of vortex logic devices \cite{16,17}.
It has also been proposed  that vortices in particular
materials can support Majorana fermions \cite{N,N2,N.2},
and that individual vortex manipulation and exchange could be used
to create certain types of quantum braiding phenomena
for quantum computing operations \cite{18,New1}.

A growing number of experiments have demonstrated individual vortex manipulation
using various techniques such as local magnetic fields \cite{19},
magnetic force tips \cite{20,21,22,23,24}, optical
methods \cite{25}, local mechanical applied stress \cite{26},
and tunneling microscope tips \cite{27,28}.
Numerous related works describe
the dynamics of individually manipulated or dragged colloidal
particles moving through glassy \cite{29,30,31,32,33,34} or crystalline
systems \cite{35,36}, where the fluctuations of the probe particle can be used to
induce local melting or to
study changes in the viscosity across an order-disorder transition.
Understanding the different kinds of dynamics associated with
particle manipulation on periodic substrates
is relevant for vortices in superconductors \cite{1,37} or
Bose-Einstein condensates \cite{38}, as well other particle based
systems with periodic substrates
such as skyrmions \cite{39}, ions on optical traps \cite{40}, colloidal particles
\cite{41,42,43,44}, and
nanofriction systems where individual atoms or molecules can be dragged with a tip \cite{45}.
In many of the previous numerical works on the local manipulation of
dragged particles, the trap used for manipulation is strong enough to permanently bind
a  single particle and drag it under a constant force.
A more accurate model of current experiments
on vortices in a type-II superconductor is
a trap of fixed strength
moving at fixed velocity that can couple to or decouple from an individual
vortex.
Vortices dragged by such a trap can either move at the average velocity of the
trap or decouple and fall away from the trap, and the trapping of multiple vortices
is also possible.

Here we consider
a trap with a finite confining force or strength
moving across a superconductor containing a periodic array of pinning sites.
As a function
of trap strength and velocity we identify five generic dynamic phases
and several subphases.
At low coupling
or high trap velocities we find a decoupled phase (I) where the trap can
only shift a vortex within a pinning site but cannot depin the vortex.
For larger coupling or smaller tip velocities,
there is an intermediate coupling phase (II)
where a single vortex can be dragged out of the pinning site but is trapped by the
next pinning site it encounters in an exchange process.
In the intermediate trapping phase (III),
vortices can be dragged over a distance of several lattice constants and additional
vortex exchange modes arise in adjacent pinning sites.
For stronger coupling, there is an intermittent multiple trapping phase (IV)
in which the trap alternates between capturing
one and two vortices,
producing telegraph noise in the trap force fluctuations.
At the strongest coupling and lowest trap velocities
we find a strong coupling phase (V) where the trap
permanently captures two vortices.
These phases are associated with distinct
signatures in the force fluctuations exerted on the moving
trap,
such as stick-slip signals associated with
vortices exiting and entering pinning sites or exchanging positions
in the trap.
We observe nonmonotonic behavior in which the trapping
effectiveness increases as the trap velocity decreases, but for the
highest trap velocities the system is
always in a decoupled phase.
We map the dynamic phases
as a function of coupling strength, trap velocity,
and the angle between the driving direction and the
pinning lattice symmetry direction.

\begin{figure}
%  \includegraphics[width=3.5in]{Fig1.pdf}
  %  \begin{minipage}{0.25\textwidth}
    \begin{minipage}{1.75in}
    \includegraphics[width=\textwidth]{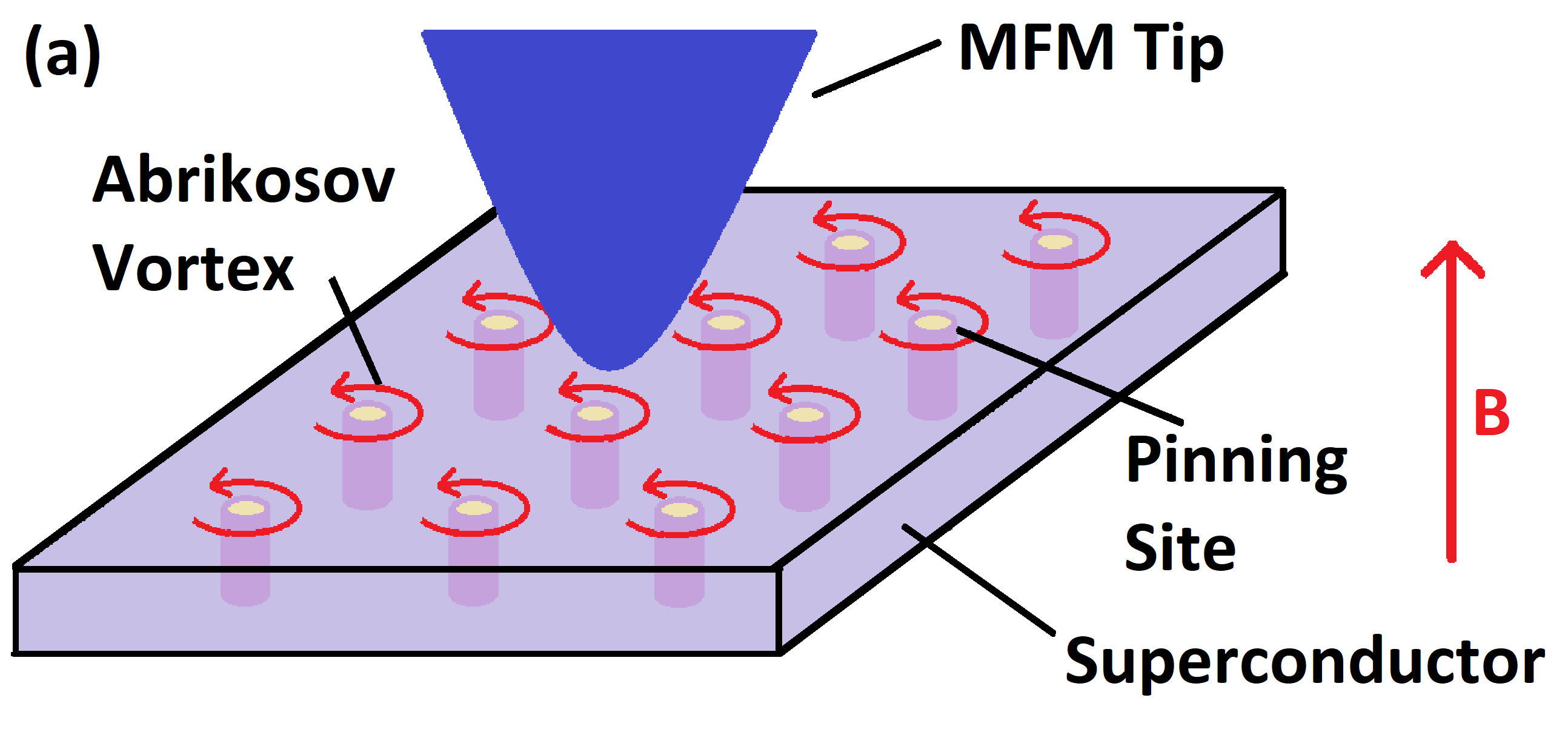}
  \end{minipage}%
%  \begin{minipage}{0.25\textwidth}
  \begin{minipage}{1.75in}
\includegraphics[width=\textwidth]{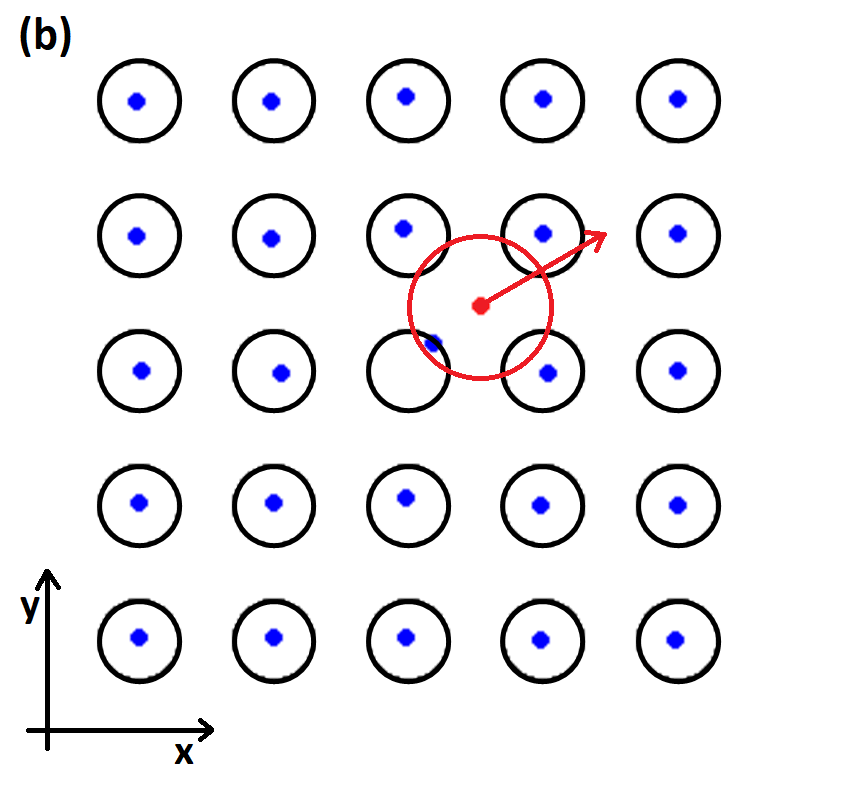}
  \end{minipage}
\caption{
  (a) A schematic of the system showing a superconducting slab containing
  a square array of artificial pinning sites (yellow) occupied by vortices (red arrows).
  The number of vortices produced by the magnetic field ${\bf B}$ applied
  perpendicular to the sample plane matches the number of pinning sites.  A magnetic
  force microscope (MFM) tip moves over the sample surface
  at velocity $v_{tr}$ and is represented by
  a finite range harmonic trap with a trapping force or strength that can be varied by
  adjusting the distance between the MFM tip and the sample.
  (b) A schematic of a $4\lambda \times 4\lambda$ subsection of the
  system.  Open black circles are pinning sites, filled blue circles are the vortices, and
  the large red circle is the trap which is
  moving at an angle of
  $\theta=30^{\circ}$
  relative  to the $x$ axis symmetry direction of the pinning array
  as indicated by the red arrow.
}
\label{fig:1}
\end{figure}

\section{Simulation and System}
We consider a two-dimensional system with periodic boundary conditions in
the $x$ and $y$-directions
containing $N_v$ vortices modeled as point particles
interacting with a square periodic pinning array.
The magnetic field applied perpendicular to the sample plane
is set to the matching field
$B=B_{\phi}$ at which the number of vortices equals the number of pinning sites.
We introduce
a trap of radius $R_{tr}$ that moves across the sample, representing a
magnetic force microscope (MFM) tip as illustrated
schematically in Fig.~\ref{fig:1}(a).
The MFM tip creates a localized
potential with a finite trapping force
that can capture one or more vortices, and it travels
at a constant velocity $V_{tr}$ at an angle
$\theta$ with respect to the $x$-axis symmetry direction of the pinning lattice.
The dynamics of vortex $i$ are determined by the overdamped
equation of motion
\begin{equation}
  \eta \frac{d{\bf r}_i}{dt}   = {\bf F}_i^{vv} + {\bf F}_i^{vp} + {\bf F}_i^{tr}.
\end{equation}
Here ${\bf r}_i$
is the position
of
vortex $i$ and
we set the damping coefficient $\eta=1$.
All forces are measured in units of $f_0=\phi_0^2/(2\pi\mu_o\lambda^3)$
where $\phi_0=h/2e$ is the flux quantum and $\lambda$ is the
London penetration depth.
The first  term on the right hand side describes the repulsive
vortex-vortex interactions,
${\bf F}_i^{vv} = \sum_{j = 1}^{N_v} K_1(r_{ij}) \hat{{\bf r}}_{ij}$,
where $r_{ij} = |{\bf r}_i - {\bf r}_j|$, ${\bf \hat{r}}_{ij} = ({\bf r}_i - {\bf r}_j)/r_{ij} $,
and $K_1$ is the
modified Bessel function of the second kind.
The pinning forces arise from a square lattice of finite range harmonic
wells,
${\bf F}_i^{vp} = -\sum_{k = 1}^{N_p} (F_p/r_p) ({\bf r}_i - {\bf r}_k^{(p)}) \Theta( r_p - |{\bf r}_i - {\bf r}_k^{(p)}| )$,
where $F_p=0.3$ is the maximum pinning force,
$r_{p} = 0.3$ is the pin radius,
${\bf r}_k^{(p)}$ is the location of the $k$-th pinning site,
and $\Theta$ is the Heaviside step
function.
The force from the moving trap ${\bf F}_i^{tr}$ has the same form as the pinning
interaction but with
a maximum trapping force of $F_{tr}$ and a trapping radius $R_{tr} = 0.5$.
The trap translates at a constant  velocity of $v_{tr}$.
We consider a $20\times20$ square pinning array at a field of $B/B_{\phi} = 1.0$,
where $B_{\phi}$ is the
matching field at which there is one vortex per pinning site.  The pinning
lattice constant is $a=1.0$ and we measure all distances in terms of
$\lambda$.
We initialize the system with each pinning site filled with a vortex.
Figure~\ref{fig:1}(b) schematically illustrates a
$4\times 4$ subsection of the sample showing
the motion of the trap, which is dragging a single vortex.
We measure the vortex displacements in and outside of the trap
as well as the time series of the force fluctuations on the moving trap.
During an individual run we translate the trap
a total distance of $D_x=300a$ in the
$x$ direction, corresponding to a total distance of $D_x/\cos(\theta)$ in the
driving direction.
Throughout this work we describe distances in terms of their projections
into the $x$ direction.

\section{Results}

\begin{figure}
\includegraphics[width=3.5in]{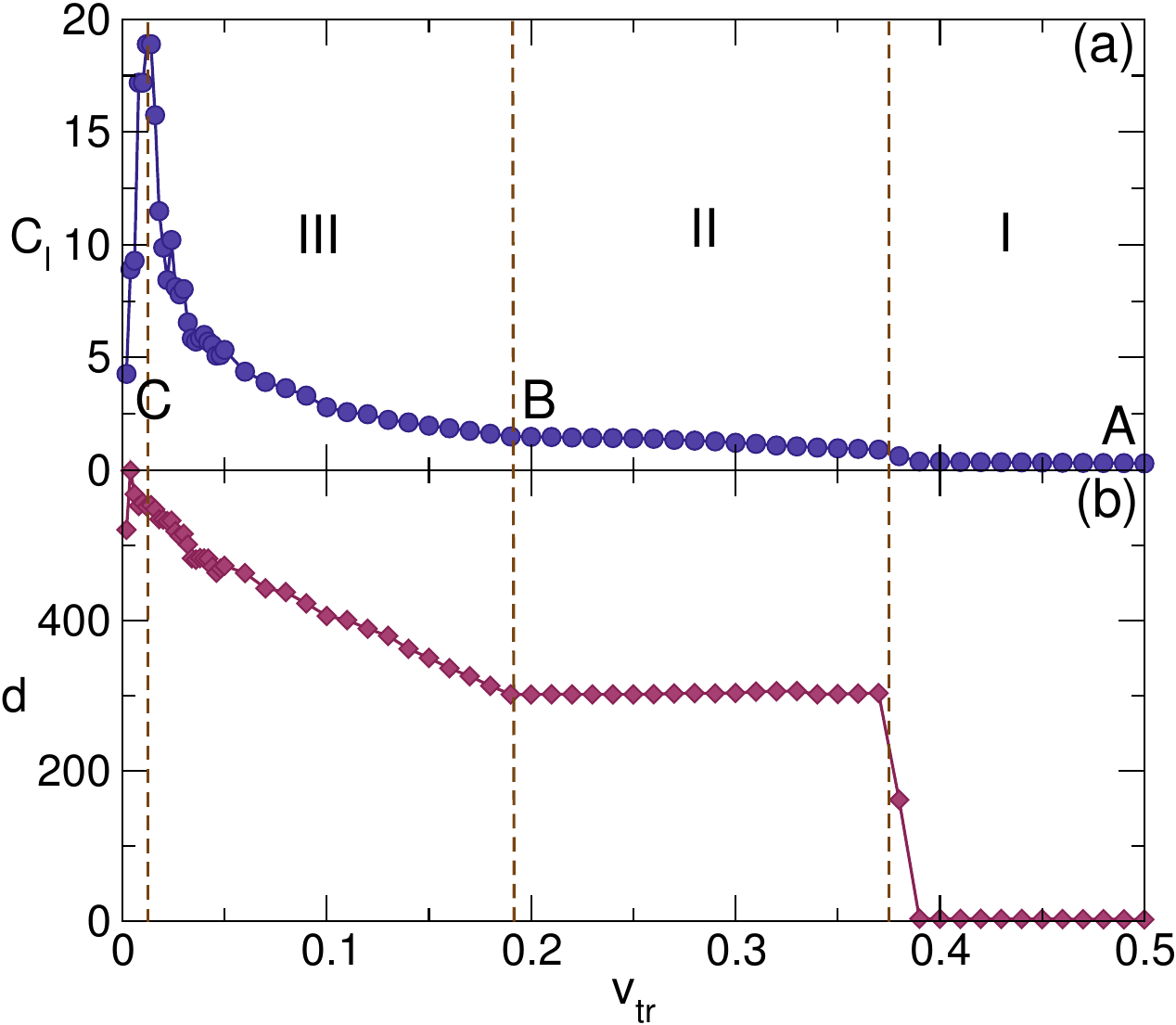}
\caption{(a) The capture length $C_l$,
  a measure of  the average distance a vortex
  is dragged by the trap,
  vs trap velocity $v_{tr}$ in a system with
  $F_{tr} = 1.0$
  where
  the trap moves at an angle of $\theta=30^{\circ}$ with respect to the
  $x$ axis of the pinning array.
  The points marked A, B, and C correspond to $v_{tr}$ values at which
  the images in Fig.~\ref{fig:3} were obtained.
  All distances have been projected into the $x$ direction.
  (b) The total displacements
  $d$ of all the vortices over a time interval during which the trap
  translates by $D_x=300a$
  vs $v_{tr}$.
  Above $v_{tr} = 0.375$, we find a transition to the decoupled phase I
  in which the trap does not drag any vortices.
  In the intermediate coupling phase II
  for $0.19 < v_{tr} <0.375$,
  an individual vortex can be dragged by the trap
  a distance of $2a$ before exchanging places with a pinned vortex.
  For $0.012<v_{tr} < 0.19$, the system is in the intermediate trapping phase III
  where vortices can be dragged
  a distance of
  several lattice constants and
  additional vortices exchange positions among the sites close to the trap.
  For $v_{tr} <  0.012$, vortices are able to escape
  more easily from the slow trap so $C_l$ drops
  even as the overall amount of displacement $d$ in the system remains high.
}
\label{fig:2}
\end{figure}

\begin{figure}
\includegraphics[width=3.5in]{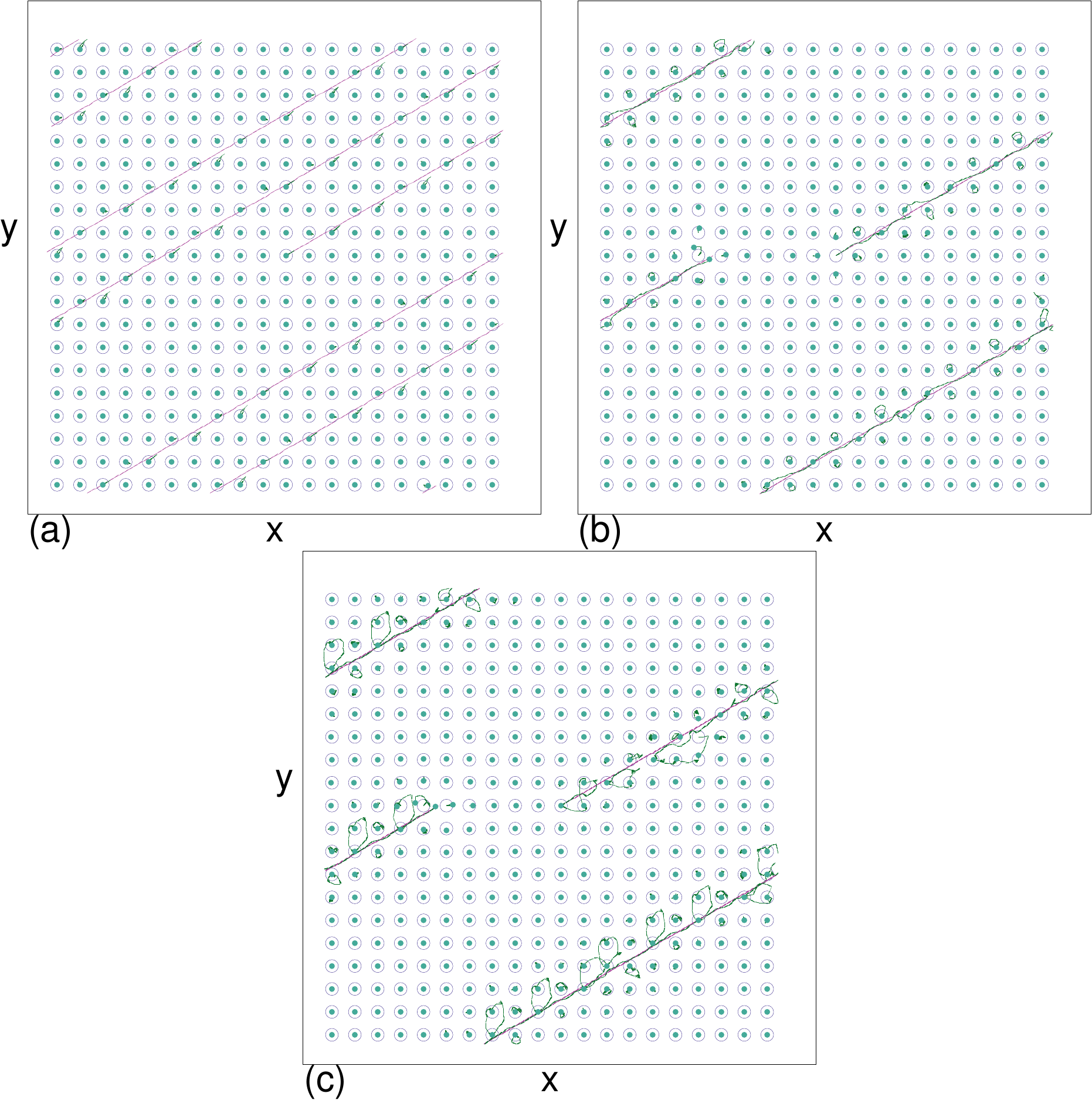}
\caption{ Vortex positions (filled circles),
  pinning site locations (open circles), tip trajectory (magenta line),
  and vortex trajectories (green lines) in the system from Fig.~\ref{fig:2}
  with $F_{tr}=1.0$, $F_p=0.3$, and $\theta=30^\circ$.
  (a) The decoupled phase I at $v_{tr} = 0.5$, marked A in Fig.~\ref{fig:2}(a),
  where all the vortices remain pinned.
  (b) The intermediate couping phase II at $v_{tr}=0.2$, marked B
  in Fig.~\ref{fig:2}(a), where individual vortices travel a distance $2a$ with the trap
  before escaping and being replaced by a new trapped vortex.
  (c) The intermediate trapping phase III at $v_{tr} = 0.02$,
  marked C in Fig.~\ref{fig:2}(a),
  where in addition to translations of the trapped vortex,
  vortices near but outside the trap move in exchange rings through
  neighboring pinning sites.
}
\label{fig:3}
\end{figure}

We define the time and location at which an individual vortex $i$ becomes
captured by the trap as ($t_{in}^i$, ${\bf r}_{in}^i$), and the corresponding time
and location at which vortex $i$ escapes from the trap as
($t_{out}^i$, ${\bf r}_{out}^i$).  We can then write the capture length
$C_l=|{\bf r}_{out}^i-{\bf r}_{in}^i|\cos(\theta)/a$ as a measure of
the distance the vortex travels inside the trap
projected into the $x$ direction and
normalized by the pinning lattice
constant $a$.
In Fig.~\ref{fig:2}(a) we plot
$C_{l}$ versus the trap velocity $v_{tr}$ for a trap with $F_{tr}=1.0$ moving at an
angle of $\theta=30^{\circ}$ with respect to the $x$ axis of the pinning array.
We observe a clear drop
in $C_{l}$  for $v_{tr} > 0.375$
when the system enters the decoupled phase I
in which the trap moves too rapidly to capture any of the pinned vortices.
In Fig.~\ref{fig:3}(a) we show the vortex and pinning site locations
along with the trajectories of the vortices and the trap
over a fixed period of time for the system in Fig.~\ref{fig:2}(a)
in the decoupled phase I
at $v_{tr} = 0.5$.
Vortices in the pinning sites wiggle a small amount as the trap passes over them
but they do not depin.

For $ 0.19 < v_{tr} < 0.375$,
we find an intermediate coupling phase II in which
the trap captures a vortex and drags it a
projected distance of approximately $2a$ to
the next pinning site along the trap trajectory, where the trapped vortex exchanges
places with the pinned vortex.
In Fig.~\ref{fig:3}(b), the vortex trajectories in phase
II at $v_{tr} = 0.2$ extend from pin to pin following the motion of the trap.
For $ v_{tr} < 0.19$ we find an intermediate trapping phase III
where individual vortices remain inside the trap for
distances greater
than $2a$ but are not permanently trapped.
Simultaneously, vortex exchange motions emerge
in the surrounding pinning sites, as illustrated
in Fig.~\ref{fig:3}(c) for $v_{tr} = 0.02$.
Figure~\ref{fig:2}(a) indicates that there
is an optimal trapping velocity $v_{tr}=0.012$ corresponding to the peak
in  $C_{l}$
where the vortex can on average be trapped for distances as large as
$18a$ before exchanging places with a pinned vortex.
For $v_{tr} < 0.012$,
$C_l$ drops dramatically when the trap velocity becomes so slow
that vortices have enough time to escape from the trap or exchange with neighboring
pinned vortices.
In contrast, for $v_{tr}>0.012$,
the trapped vortex
can remain trapped since it does not have enough time to exchange with
another vortex.  As $v_{tr}$ increases above 0.012, $C_l$ drops as
the trapped vortex experiences larger displacements
until the system reaches the II-III transition where
the trapped vortex always exchanges with a pinned vortex.

To measure the
global effect of the trap,
in Fig.~\ref{fig:2}(b) we plot the scaled net total
projected displacement $d$ of all the vortices
$d = a^{-1}\sum^{N_{v}}_{i= 0}|({\bf r}^{i}(t_0+\tau) - {\bf r}^{i}(t_{0}))\cdot {\bf \hat{x}}|$
versus $v_{tr}$, where
$\tau=D/(v_{tr}\cos(\theta))$ is the time required for the trap to
translate a projected distance of $D=300a$.
Above the I-II transition at $v_{tr}=0.375$, $d$ drops to zero.
In the intermediate coupling phase II, the trap is never empty,
and there is a plateau with $d=300$ 
throughout the phase II region of 
$0.19 < v_{tr} < 0.375$.
No individual vortex travels this distance with the trap; instead,
as shown in Fig.~\ref{fig:2}(a),
vortices
translate an average distance
of $C_l=2a$
before encountering a pinning site
and exchanging with the pinned vortex.
The surrounding vortices remain pinned and do not contribute to $d$.
In the intermediate trapping phase III
for $v_{tr} < 0.19$,
$d$ increases
with decreasing $v_{tr}$ as vortices surrounding the trap
begin to depin from the pinning sites
and undergo rotational exchange motions of the type illustrated
in Fig.~\ref{fig:3}(c) at $v_{tr} = 0.02$.

\begin{figure}
\includegraphics[width=3.5in]{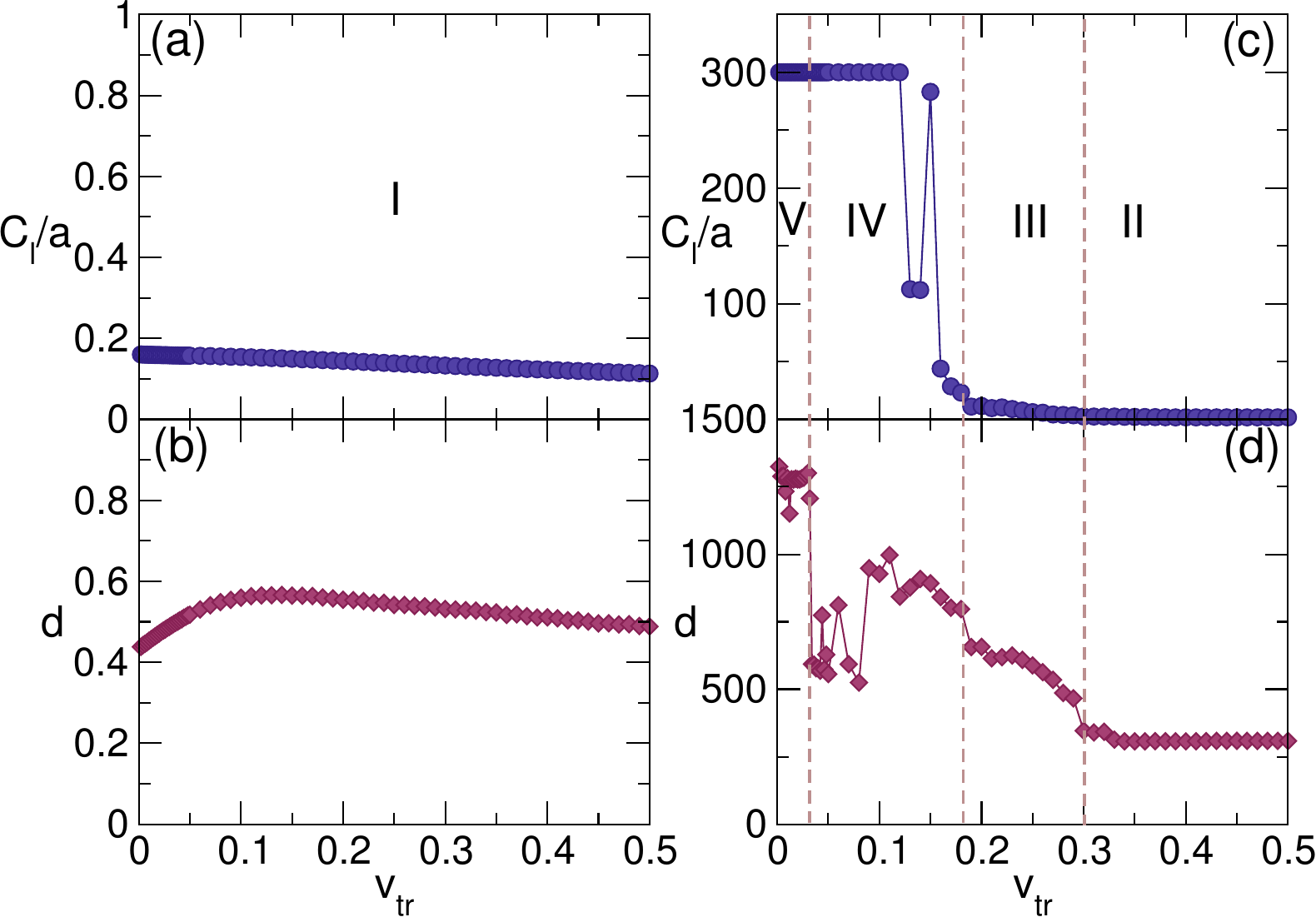}
\caption{ (a) $C_{l}$ vs $v_{tr}$ and
  (b) $d$ vs $v_{tr}$ for the $\theta=30^\circ$ system with
  a decreased trap strength of
  $F_{tr} = 0.5$.
  The motion is always
  in the decoupled phase I.
  (c) $C_{l}/a$ vs $v_{tr}$ and
  (d) $d$ vs $v_{tr}$ in the same system for
  a strong trap with $F_{tr} = 1.8$,
  where dashed lines indicate the boundaries of phases
  II, III, IV, and V.
  In the intermittent multiple trapping phase IV, the trap intermittently
  captures two vortices, and in the strongly coupled
  phase V, the trap always captures two vortices.
}
\label{fig:4}
\end{figure}

\begin{figure}
\includegraphics[width=3.5in]{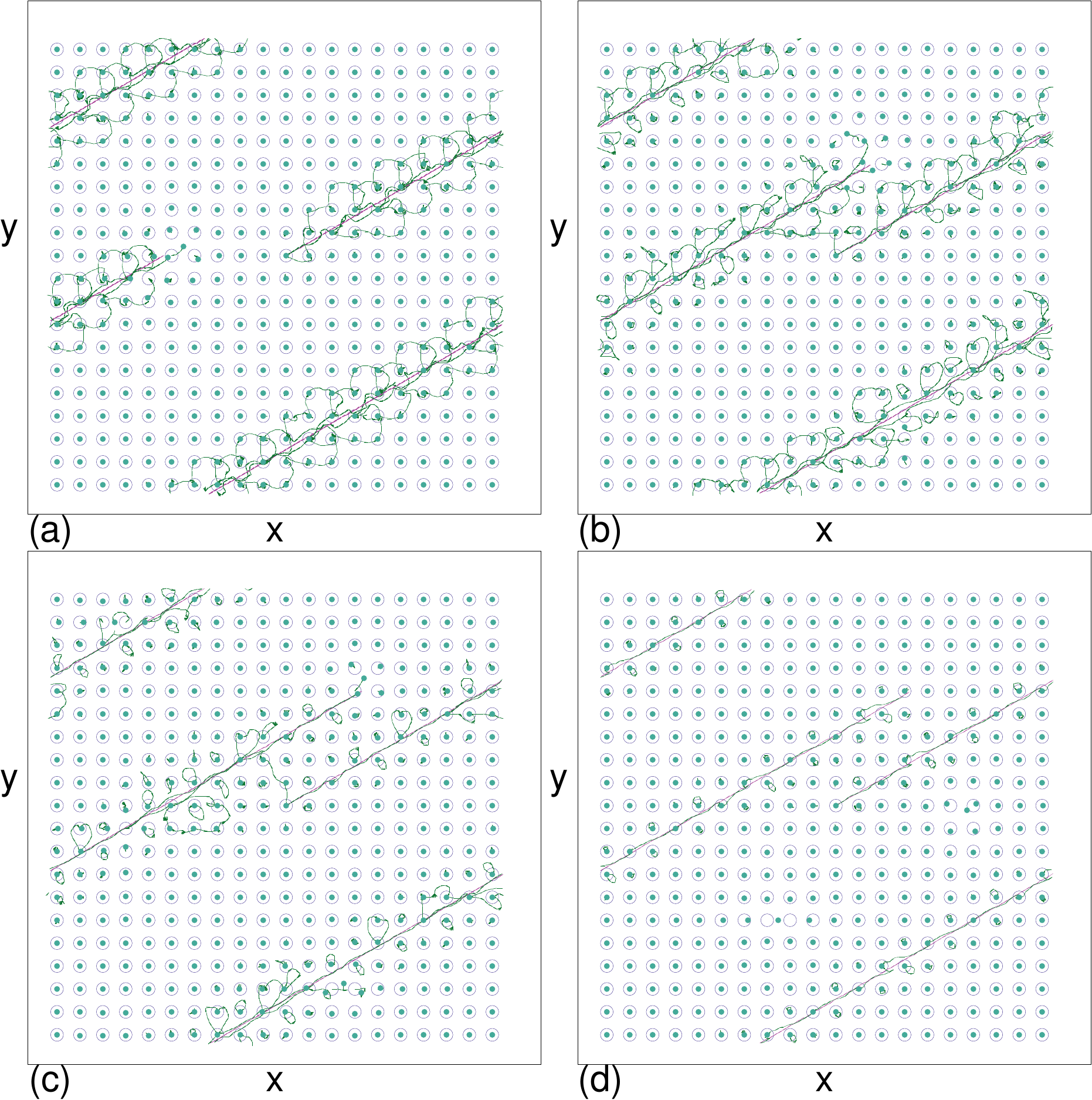}
\caption{
  Vortex positions (filled circles), pinning site locations (open circles), tip trajectory
  (magenta line), and vortex trajectories (green lines)
  for the $F_{tr}=1.8$ system in Fig.~\ref{fig:4}(c,d).
  (a) At $v_{tr} = 0.02$ in phase V,
  the trap always contains two vortices and correlated ringlike
  exchanges of vortices occur in the surrounding regions.
  As the trap moves, the two trapped vortices dislodge three pinned vortices, and
  one of these vortices jumps into the empty pinning site immediately
  behind the moving trap.
  (b) At $v_{tr} = 0.12$ in phase IV, the trap alternates between capturing one and two vortices.
  (c) Disordered flow in phase III at $v_{tr}=0.25$, with
  a much weaker perturbation of the surrounding pinned vortices.
  (d) At $v_{tr}= 0.5$ in phase II, a vortex only travels a short distance with the trap
  before exchanging with a pinned vortex.
}
\label{fig:5}
\end{figure}

We find transitions among the different phases as a function of
trap strength $F_{tr}$ as well as trap velocity.
Figure~\ref{fig:4}(a,b) shows $C_{l}$ and $d$
versus $v_{tr}$ for driving at $\theta= 30^{\circ}$
in the same system from Fig.~\ref{fig:2} with a
smaller $F_{tr} = 0.5$.
Both $C_{l}$ and $d$ are less than one, and
the system remains in the decoupled phase
I for all values of $v_{tr}$.
When we instead consider a larger trap strength
$F_{tr} = 1.8$,
Fig.~\ref{fig:4}(c,d) shows that
phase II appears
for $v_{tr} > 0.3$, while
for $v_{tr} < 0.03$ the system is in phase V and the trap always contains
two vortices.
In Fig.~\ref{fig:5}(a) we
illustrate the vortex trajectories in phase V at
$v_{tr} = 0.02$,
where a multi-vortex exchange process occurs in the vortices adjacent to the trap.
The two trapped vortices produce a repulsion that is strong enough to depin
the vortex in the pin traversed by the trap along with those in
a pair of neighboring pins on
either side of the trap.  The three depinned vortices form a cascading loop of
reoccupancy, and one of them moves to occupy the pinning site
behind the trap that was previously vacated.
For $0.018 < v_{tr} <  0.3$ we find the intermediate trapping phase III,
while for $ 0.03 < v_{tr} < 0.018$
an intermittent multiple trapping phase IV occurs in which
trap alternates between capturing one or two vortices.
Figures~\ref{fig:5}(b) and (c) illustrate typical phase IV and phase III trajectories,
respectively.
Phase IV contains several subregimes.
When $v_{tr}$ is close to 0.1,
the trap permanently captures one vortex and exchanges a second vortex
with each pinning site it passes,
while at larger $v_{tr}$,
both trapped vortices exchange places with vortices in the pinning sites as the
trap moves.

\begin{figure}
\includegraphics[width=3.5in]{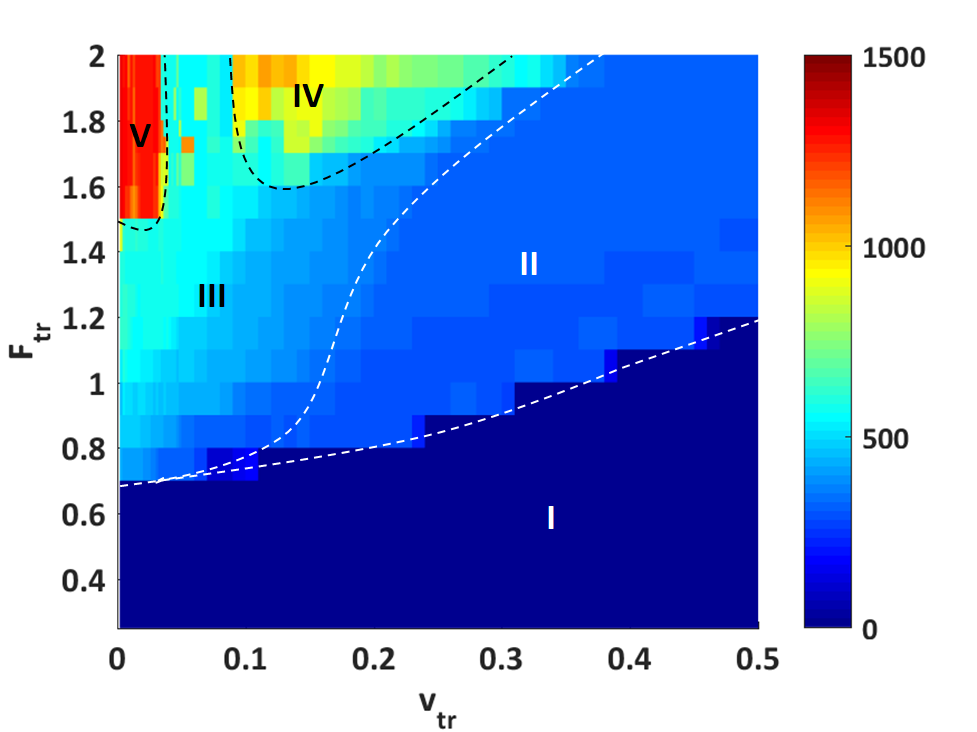}
\caption{Heat map of the total displacements $d$ as a function of
  $F_{tr}$ vs $v_{tr}$ for driving
  at $\theta = 30^\circ$. Dashed lines are guides to the eye indicating
  the locations of the different phases: I (decoupled), II (intermediate coupling),
  III (intermediate trapping), IV (intermittent multiple trapping) and V (strongly coupled).
}
\label{fig:6}
\end{figure}

\begin{figure}
  \begin{minipage}{3.5in}
    \includegraphics[width=\columnwidth]{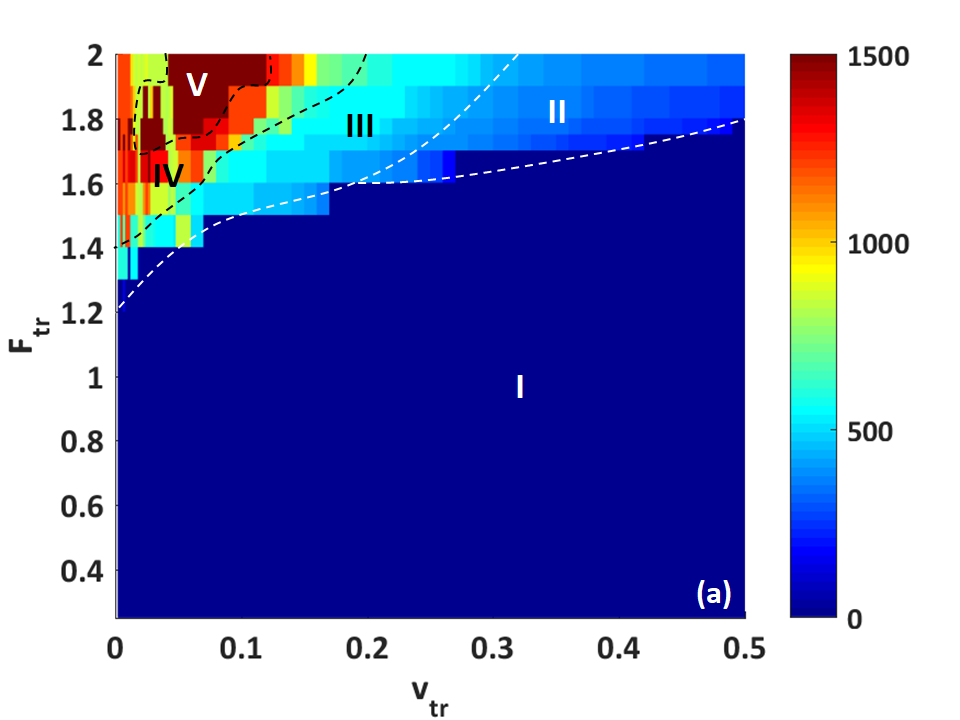}
    \includegraphics[width=\columnwidth]{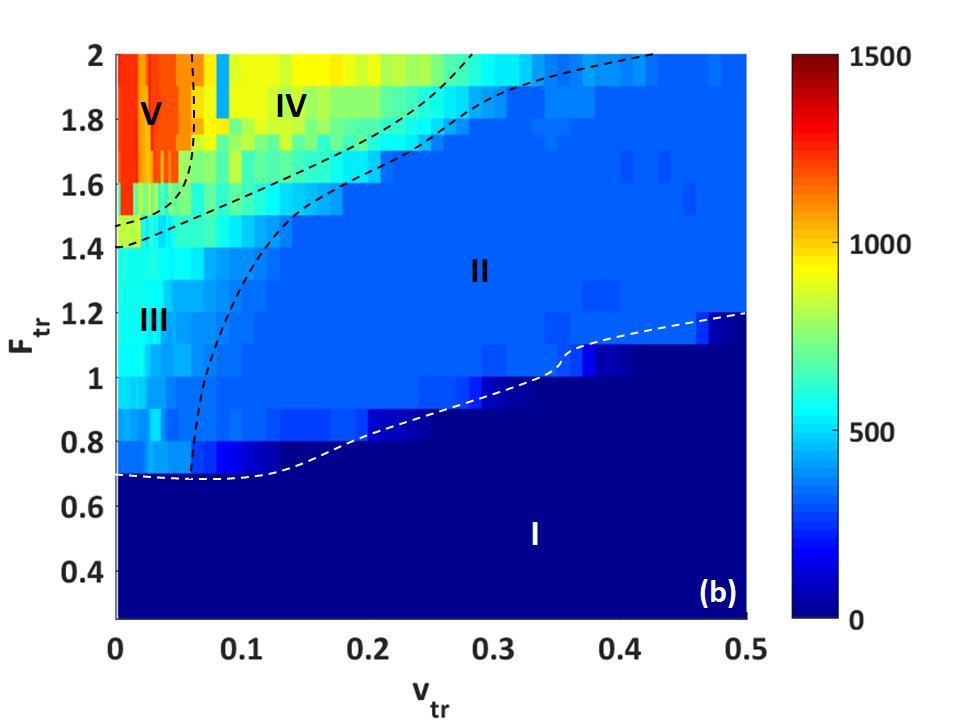}
    \includegraphics[width=\columnwidth]{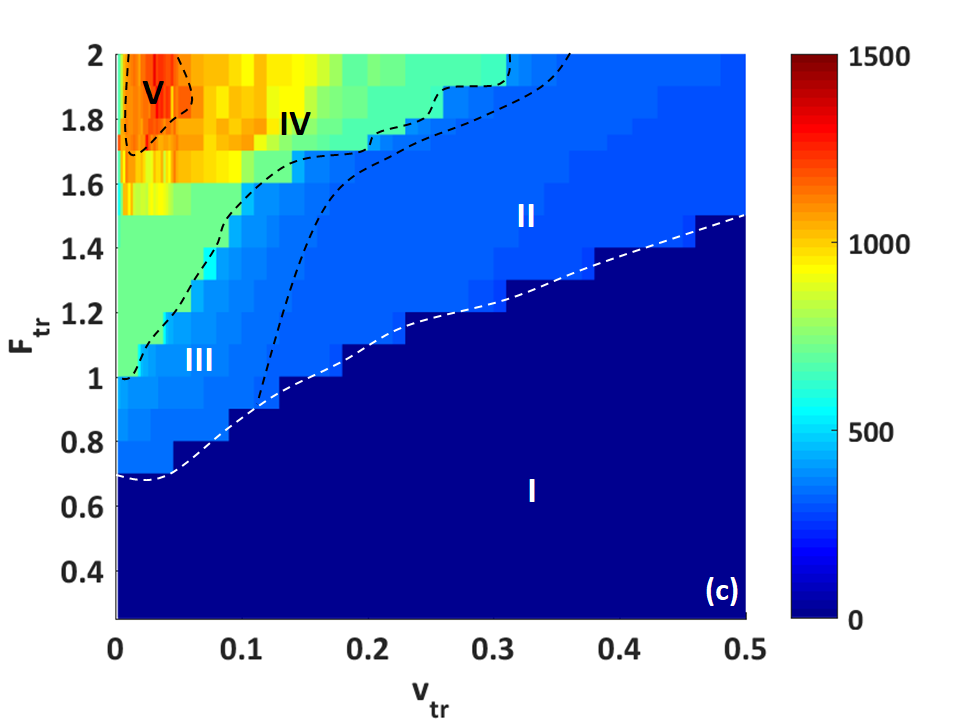}
  \end{minipage}
  \caption{ Heat maps of $d$ as a function of $F_{tr}$ vs $v_{tr}$ for driving at
    (a) $\theta = 0^\circ$,
    (b) $\theta=15^\circ$,
    and (c) $\theta=45^\circ$. The locations of phases I to V are marked by lines that serve
    as guides to the eye.
}
\label{fig:7}
\end{figure}

\begin{figure}
\includegraphics[width=3.5in]{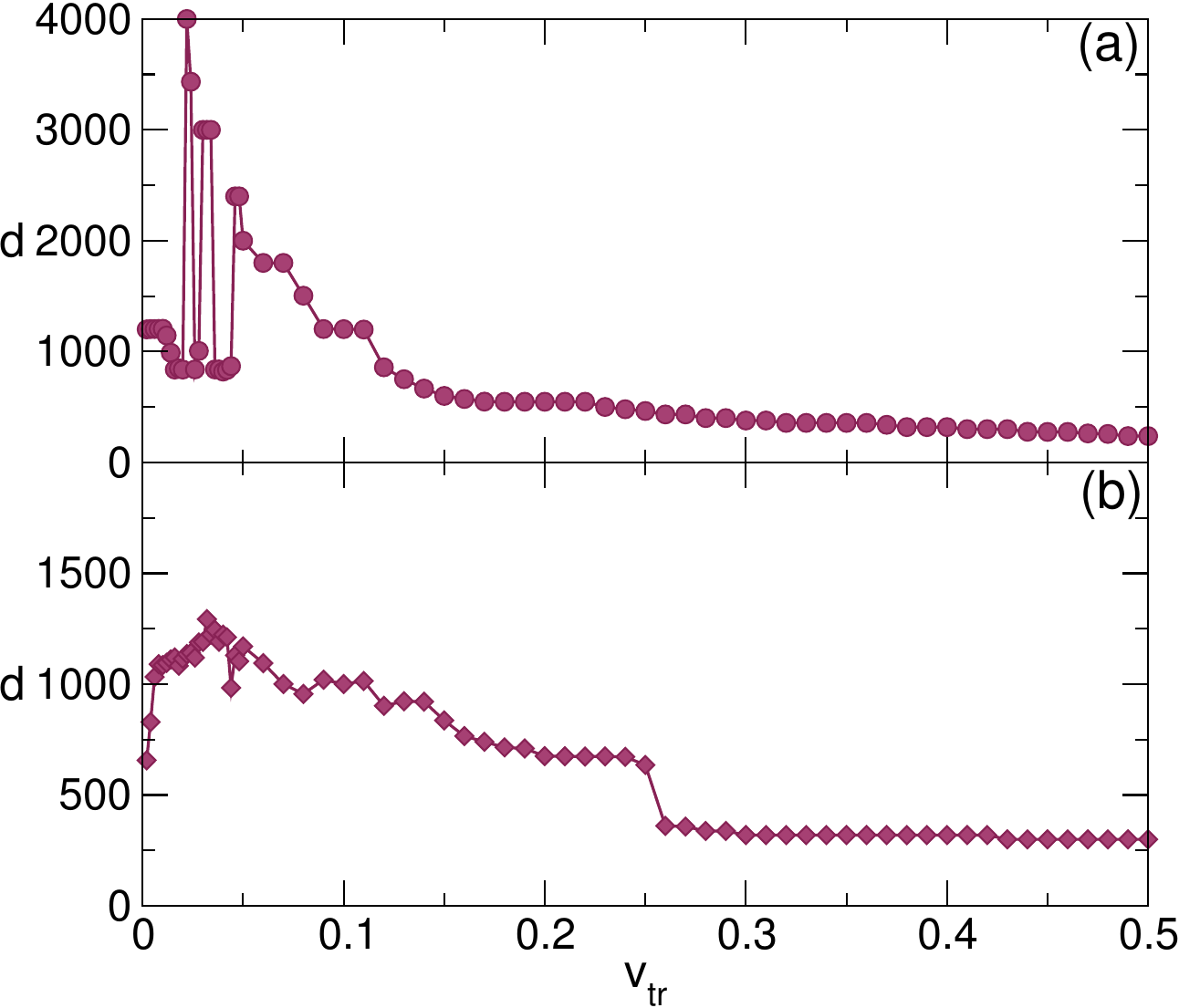}
\caption{ $d$ vs $v_{tr}$ at $F_{tr}=1.8$ for
  (a) $\theta = 0^\circ$
  and
  (b) $\theta = 45^\circ$.
}
\label{fig:8}
\end{figure}

In Fig.~\ref{fig:6} we plot a heat map of the total displacements $d$
as a function of
trap strength $F_{tr}$ versus trap velocity $v_{tr}$ for
a driving angle of $\theta=30^\circ$ in which we highlight the locations of
phases I through V.
For $F_{tr} < 0.75$,
the system is in the decoupled phase I.
The effect of changing the angle of drive on $d$ appears in the
$F_{tr}$ versus $v_{tr}$ heat maps in
Fig.~\ref{fig:7}(a,b,c)
for $\theta = 0^{\circ}$, $15^{\circ}$, and $45^{\circ}$,
which trace the evolution of the five different phases.
In each case, the transition lines generally shift
to higher values of $F_{tr}$ with increasing $v_{tr}$.
For $\theta = 0^{\circ}$ in Fig.~\ref{fig:7}(a), the trap does not
start dragging vortices out of the pinning sites until
$F_{tr} > 1.25$, and we observe a variety of additional subphases that are not
present at larger $\theta$.
Previous work for
vortices driven over square periodic pinning
arrays at $\theta = 0^\circ$
showed  a series of distinct dynamical
phases associated with positive or negative jumps
in the velocity-force curves \cite{7,8,46,47,48},
and the dynamics we observe in Fig.~\ref{fig:7}(a) is consistent with
this type of behavior.
In Fig.~\ref{fig:8}(a) we plot $d$ versus $v_{tr}$
for the $\theta = 0^\circ$ sample
at $F_{tr} = 1.8$,
where we find numerous jumps at small $v_{tr}$,
while in Fig.~\ref{fig:8}(b), $d$ versus $v_{tr}$
at
$F_{tr}=1.8$
and $\theta=45^\circ$
has a smoother behavior at small $v_{tr}$ and a step marking the II-III transition
at $v_{tr}=0.25$.

In order for a system
in the decoupled phase I to reach the intermediate coupling phase II, in which the
trap is able to drag vortices from pin to pin, a vortex must remain inside
the trap during the entire time required for the trap to traverse the distance $d_p$
between pins.
The trap can interact with the finite size pin
once it is 
within a distance $r_{p}$
of the pin, giving
$d_p=(a-r_{p})/C$, where $C=1$ for $\theta=0^\circ$ and
$C=\sin{\theta}$ for $0^\circ < \theta < 90^\circ$.
The time required to travel this distance is $\delta t_1=d_p/v_{tr}$.
During the entire interval $\delta t_1$, the vortex experiences two
competing forces: the trapping force $F_{tr}$ that pulls the vortex
toward the center of the trap, and a
harmonic restoring force $F_{r}$
from the background vortex lattice that pulls
the vortex toward its equilibrium lattice position.
There is also a velocity-dependent
drag term $\eta v_{tr}$ that
represents an effective
viscosity produced by dynamic rearrangements of the background
vortex lattice.
During the time interval $\delta t_2=2r_p/Cv_{tr}$ spent traversing a pin
the vortex is also subjected to the pinning force $F_p$.
The maximum displacement the
vortex can experience under these forces while traveling between pinning sites
is then
$\Delta r=\delta t_1(F_{tr}-F_r-\eta v_{tr})/\eta-\delta t_2F_p/\eta$.
When $\Delta r \leq r_p$, the vortex remains inside the pin and cannot be
captured by the trap, placing the system in phase I,
but when $\Delta r >r_{p}$, the trap can pull the vortex out of the pin and
carry it to the next pin,
and the system enters phase II.
Thus, we predict that the I-II transition should occur when
$r_{p}=(a-r_{p})(F_{tr}-F_r-\eta v_{tr})/C \eta v_{tr}-2r_pF_p/C\eta v_{tr}$, which can be written as
\begin{equation}
   F_{tr}=\frac{Cr_{p}\eta v_{tr}+2r_pF_p}{a-r_{p}}+F_r+\eta v_{tr}.
\end{equation}

To obtain an estimate of the value of $F_r$, first consider the $\theta=0^\circ$ case.
As the trap passes over a pinned vortex,
which we call vortex A, it tends to shift  vortex A away from the
center of the pin toward the leading or trailing edge of the trap, depending on whether
the trap is just arriving at the pinning site or just leaving it.  Vortex-vortex interactions
cause a similar shift in the position of the neighboring pinned vortices due to the high
symmetry of the $\theta=0^\circ$ motion.  Vortex A can
thus
be regarded as sitting inside three potential wells: the trap potential, the pin potential,
and a harmonic potential produced by the surrounding vortex lattice that provides the
restoring force $F_r$.  When $\theta=0^\circ$, this restoring force is dominated by
interactions from the four closest vortices that are a distance $a$ from vortex A.
The magnitude of the restoring force, to lowest order, can be estimated as
$F_r=[K_1(a)-K_1(a+\delta r)]/\delta r$.  For $\delta r=0.005$, we obtain
$F_r=1.017$ for $\theta=0^\circ$.
The motion becomes less symmetric once $\theta>0^\circ$, and the four closest
vortices no longer contribute to an effective harmonic confining potential due to
their asymmetric distortions.
Instead, the next four closest vortices
that are a distance $\sqrt{2}a$ from vortex A
maintain sufficient symmetry in their motion to produce a weaker harmonic confining
force.  We thus obtain
$F_r=[K_1(\sqrt{2}a)-K_1(\sqrt{2}a+\delta r)]/\delta r$, giving
$F_r=0.459$ for $\delta r=0.005$
and $0^\circ<\theta<90^\circ$.

From Eq.~2, the predicted $v_{tr}=0$ intercepts are $F_{tr}=1.27$ for $\theta=0^\circ$ and
$F_{tr}=0.716$ for $0^\circ<\theta<90^\circ$ due to the different values of $F_r$ in the two
cases.
At $v_{tr}=0.5$, the I-II transition line is predicted to fall at
$F_{tr}=1.99$ for $\theta=0$, $F_{tr}=1.27$ for $\theta=15^{\circ}$,
$F_{tr}=1.32$ for $\theta=30^\circ$,
and $F_{tr}=1.36$ for $\theta=45^\circ$.
These predictions are in general agreement
with the results in Figs.~\ref{fig:6} and \ref{fig:7},
but are not exact
due to our neglect of higher-order 
contributions to $F_r$ and the influence of the nonsymmetric arrangement of the
closest trapped vortices.
For the $\theta=30^\circ$
system in Fig.~\ref{fig:4}(c,d) with $F_{tr}=1.8$, Eq.~2 predicts
a I-II transition at $v_{tr}=0.9$, outside the range of values we
consider, while there is no value of $v_{tr}$ satisfying Eq.~2 when
$F_{tr}=0.5$ and $\theta=30^\circ$ since the $v_{tr}$
intercept falls at $F_{tr}=0.716$, indicating that the system in
Fig.~\ref{fig:4}(a,b) should never undergo a I-II transition.
For $\theta = 45^{\circ}$ in Fig.~\ref{fig:7}(c),
the I-II line separating the decoupled and partially coupled phases
is slightly steeper than in the $\theta=30^\circ$ and $\theta=15^\circ$ samples,
as expected from Eq.~2 due to the larger value of $\sin{\theta}$.
The $\theta=0^\circ$ and $F_{tr}=1.8$ sample in Fig.~\ref{fig:8}(a) is
predicted to have a I-II transition at $v_{tr}=0.37$,
while for the $\theta=45^\circ$ and $F_{tr}=1.8$ sample the I-II transition
is predicted to fall at $v_{tr}=0.83$.

\begin{figure}
\includegraphics[width=3.5in]{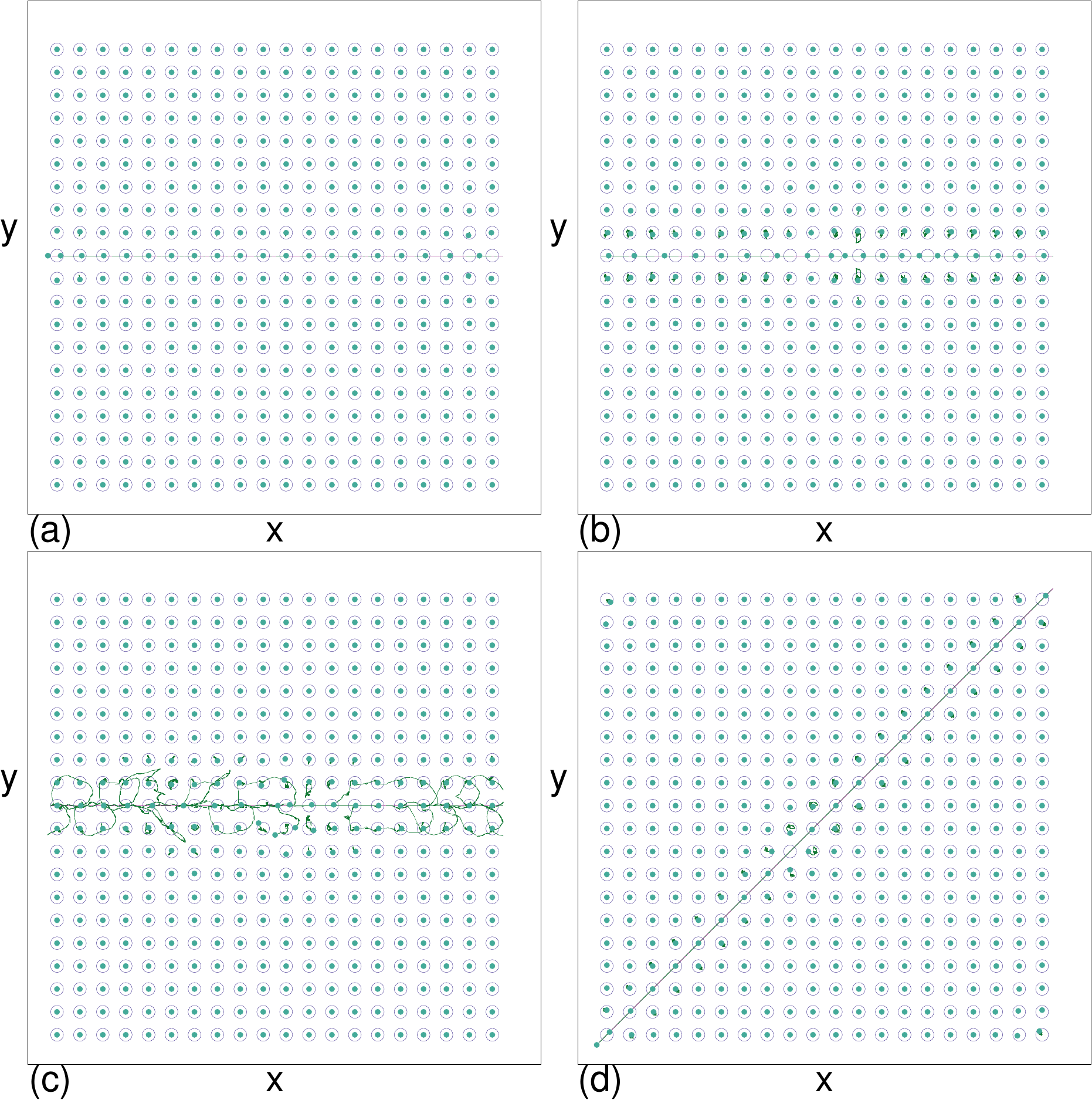}
\caption{ Vortex positions (filled circles),
  pinning site locations (open circles), tip trajectory (magenta line), and
  vortex trajectories (green lines)
  for a sample with $F_{tr}=1.8$ and (a-c) $\theta=0^\circ$.
  (a) Phase II at $v_{tr}= 0.35$, where there is little distortion of the background.
  (b) Phase III at $v_{tr} = 0.1$, where the amount of distortion of the surrounding
  vortices has increased.
  (c) Phase IV at $v_{tr} = 0.02$,
  where the multiple dragged vortices induce plastic motion in the
  surrounding vortices.
  (d) The phase II motion at $v_{tr}=0.35$ in a sample with
  $\theta=45^\circ$.
}
\label{fig:9}
\end{figure}

In Fig.~\ref{fig:9}(a) we show the vortex and trap trajectories in phase II for
a sample with $F_{tr}=1.8$ and $\theta=0^\circ$ at
$v_{tr} = 0.35$.
Individual vortices are trapped over a distance of one lattice constant,
moving along a one-dimensional path defined by the trap trajectory and inducing
few to no perturbations in the surrounding vortices before exchanging
positions with the next pinned vortex along the path of the trap.
At $v_{tr}=0.1$ in Fig.~\ref{fig:9}(b),
the perturbations to the surrounding vortices are stronger,
while at $v_{tr}=0.02$ in
Fig.~\ref{fig:9}(c),
there is continuous plastic mixing of the vortices
in the two rows of pins on either side of the trap trajectory.
For driving along $\theta=45^\circ$,
Fig.~\ref{fig:9}(d) shows that in phase II at
$F_{tr} = 1.8$ and $v_{tr} = 0.35$,
motion occurs along the diagonal with some distortions of the
vortices in the adjacent pinning sites.

\section{Force Fluctuations}

\begin{figure}
\includegraphics[width=3.5in]{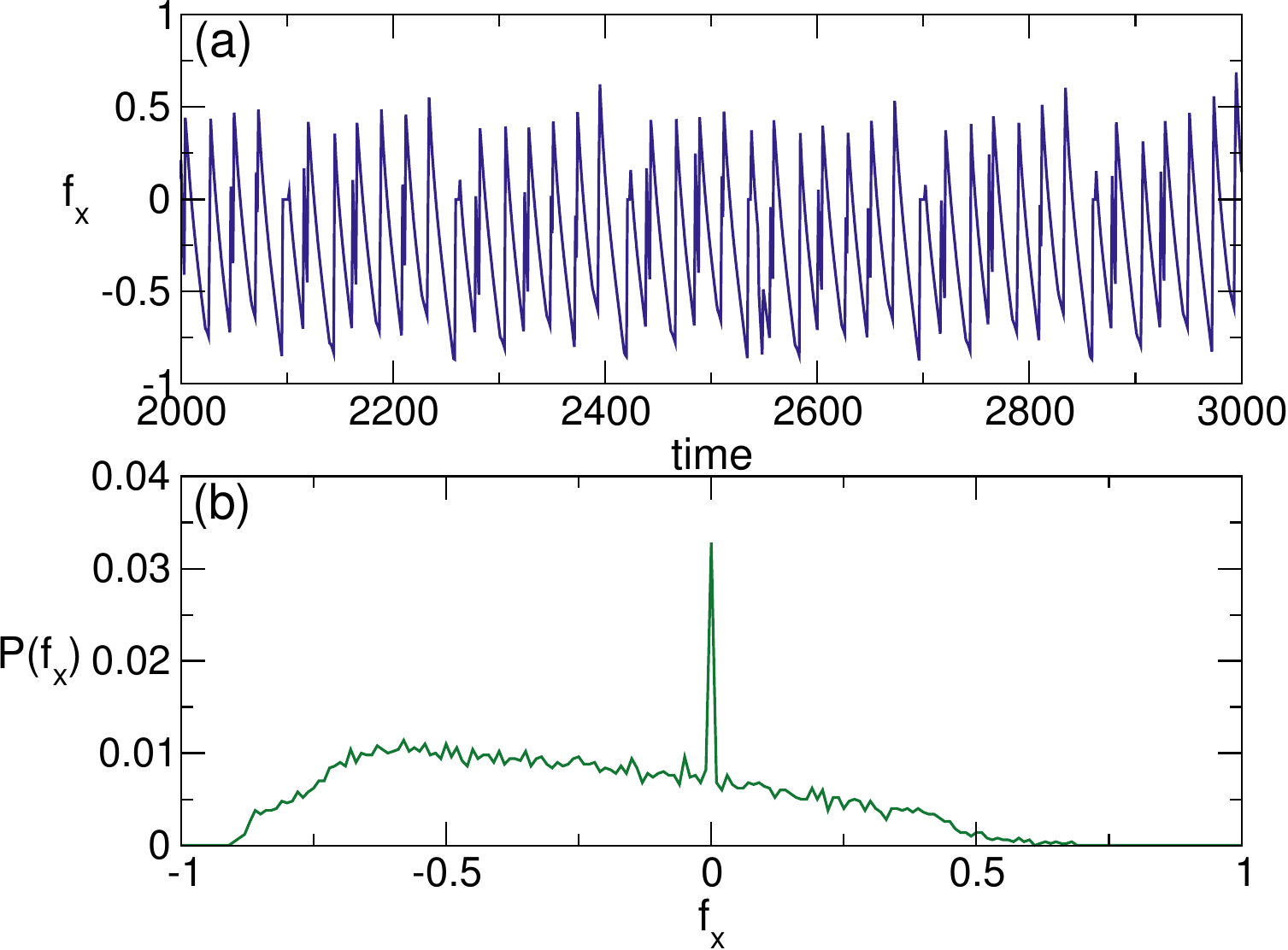}
\caption{ (a) A representative plot
  of the time series of the $x$ direction forces
  $f_x$
  experienced by the moving trap in phase I at
  $\theta = 30^{\circ}$, $F_{tr} = 1.0$, and $v_{tr} = 0.5$.
  (b) The corresponding distribution function $P(f_x)$.
  The time intervals when the trap does not contain a vortex produce
  the peak at $f_x = 0$.
}
\label{fig:10}
\end{figure}

We next examine the time series of the
$x$ direction forces $f_x$ experienced by the trap as it moves
in the different phases.
In Fig.~\ref{fig:10}(a)
we
plot a representative time series of $f_x$ for the
$\theta=30^\circ$ system from Fig.~\ref{fig:2}(a)
and Fig.~\ref{fig:3}(a)
in the decoupled  phase I
at
$F_{tr} = 1.0$ and $v_{tr} = 0.5$.
We find a pronounced stick-slip character in $f_x$ with a strong asymmetry of
sudden increases and gradual decreases.
The slow drops in $f_x$ occur
when the moving trap is dragging
a vortex inside a pinning site and the
force from  the pinning site is resisting the pull of the trap,
while the rapid increases correspond to intervals
when the vortex decouples from the trap and drops back into the pinning site.
Figure~\ref{fig:10}(b) shows that
the probability distribution function $P(f_{x})$
has a spike at $f_{x} = 0$ produced by
the time periods during which
there is no vortex inside the trap.
There is a local maximum in $P(f_x)$ near $f_x \approx -0.6$, the value of
the $x$ component of the average decoupling force $F_{\rm dc}$ at which
the vortex escapes from the trap.
At decoupling, the vortex is at the edge of the pinning site, where it experiences
a force of magnitude $F_p$,  
and it is a distance $r_p$ from the center of the trap,
where it is subjected to
a force of magnitude $F_{tr}(r_p/R_{tr})$.
There is also a drag force contribution of $C\eta v_{tr}$ from the background of
pinned vortices.
This gives a decoupling force of
$F_{\rm dc}=-F_{p}-F_{tr}(r_p/R_{tr})+C\eta v_{tr}$,
which for
$\theta=30^\circ$,  $F_{p} = 0.3$,
and $v_{tr} = 0.5$
gives $F_{\rm dc} = -0.65$,
in agreement with the location of the local maximum in $P(f_x)$.
As $v_{tr}$ increases, we find that the local maximum in $P(f_x)$ shifts
to lower absolute values of $f_x$.

\begin{figure}
\includegraphics[width=3.5in]{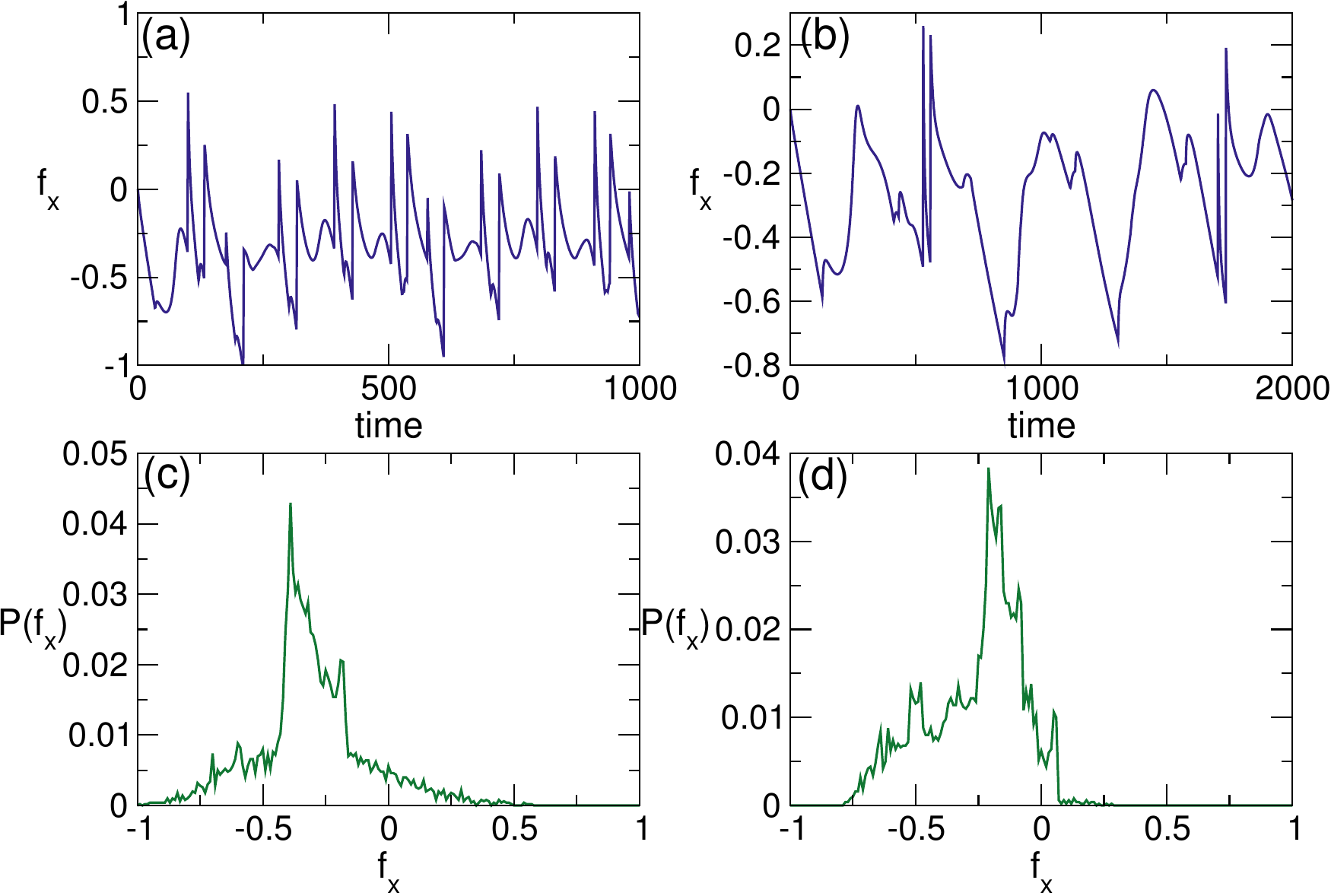}
\caption{
  (a) A representative segment of $f_x(t)$ in phase II at $v_{tr}=0.2$ for a sample with
  $\theta = 30^{\circ}$ and $F_{tr} = 1.0$.
  (b) The corresponding $P(f_x)$.
  There is no peak at $f_x=0$ since the trap always contains a vortex.
  (c) $f_x(t)$ in the same sample in phase III at $v_{tr}=0.05$.
  (d) The corresponding $P(f_x)$ contains
  additional peaks
  produced by additional modes of motion.
}
\label{fig:11}
\end{figure}

In Fig.~\ref{fig:11}(a,b) we plot $f_x(t)$ and $P(f_x)$ in phase II at $v_{tr}=0.2$ for
a sample with
$\theta = 30^\circ$ and
$F_{tr} = 1.0$.
There is no longer a peak in $P(f_x)$ at
$f_x = 0$
since the trap always contains one vortex.
We find
a periodic signal in $f_x(t)$ containing
both stick-slip features and
additional smoother oscillations
between pairs of force spikes.
The force spike pairs arise when the trap captures a new vortex or drops
a trapped vortex.
Since a trap with $F_{tr}=1.0$ is not strong enough to confine two vortices,
every time the trap captures a vortex it sheds
the previously captured vortex.
The process of bringing a trapped vortex close to a pinned vortex, followed by
capture of the pinned vortex, produces a peak in $P(f_x)$ at $f_x=-0.4$.
The smooth oscillations occurring on a
longer time scale correspond to the transport of a vortex between pinning sites
by the trap, since at $\theta=30^\circ$
the trap passes over a pinning site in every other column of the pinning array.
When the trap passes between two pinned vortices at a distance $a/2$,
the trapped vortex must cross an energy barrier generated by the
repulsive vortex-vortex forces, giving a second peak in $P(f_x)$ at $f_x=-0.2$.

In Fig.~\ref{fig:11}(c,d) we show $f_x(t)$ and $P(f_x)$ for $v_{tr}=0.05$ in
phase III  for the
$\theta=30^\circ$ and $F_{tr}=1.0$ system from Fig.~\ref{fig:11}(a,b).
At this low trap velocity, the trapped vortex produces a larger perturbation of the
surrounding vortices
as it moves, 
resulting in the appearance of additional peaks in $P(f_x)$.
The highest peak in $P(f_x)$ at $f_x=0.45$
results when the strongly
trapped vortex passes through a pinning site and pushes the pinned vortex out of
its way without escaping from the trap.

\begin{figure}
\includegraphics[width=3.5in]{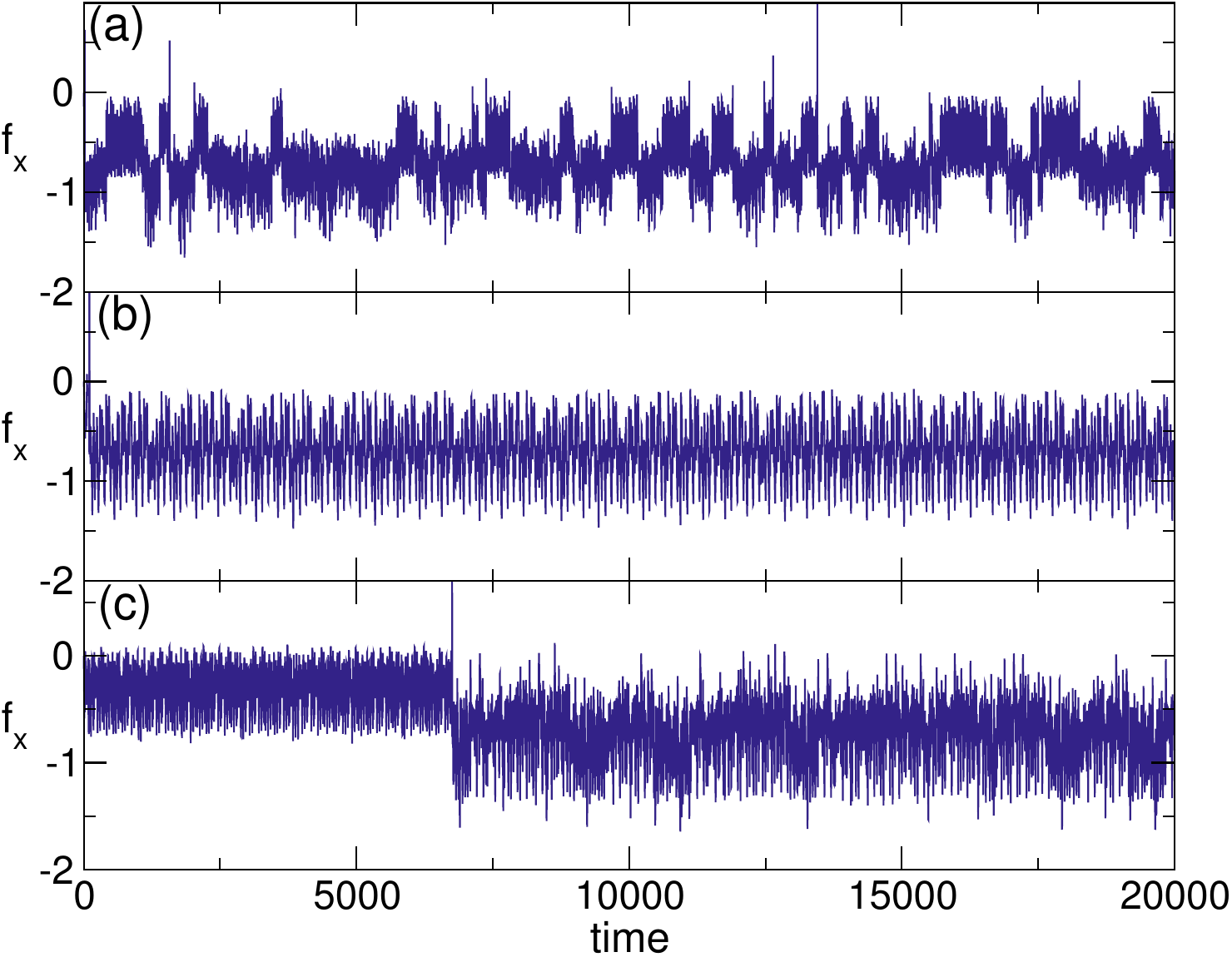}
\caption{
  (a)
  $f_x(t)$ for a sample with
  $\theta = 30^{\circ}$ and $F_{tr}=1.8$ at
  $v_{tr} = 0.12$
  in phase IV.  The time series has a telegraph noise characteristic in which
  the two values are produced when the trap
  alternates between dragging one (higher $f_x$) or two (lower $f_x$) vortices.
  (b) In phase V at
  $v_{tr} = 0.02$, the trap always captures two vortices
  and the telegraph noise is lost.
  (c) In phase IV at $v_{tr}=0.048$, there is a transient
  signature when the trap initially drags one vortex but then captures a
  second vortex, producing a clearly visible jump in $f_x$.
}
\label{fig:12}
\end{figure}

In phase IV,  illustrated for a sample with
$\theta=30^\circ$ and $F_{tr}=1.8$  at
$v_{tr} =0.12$ in Fig.~\ref{fig:12}(a),
$f_x(t)$ shows a strong telegraph noise signal in which
two states arise when the trap alternates between dragging
one or two vortices.
In phase V at $v_{tr}=0.02$,
Fig.~\ref{fig:12}(b) indicates that the telegraph noise in $f_x(t)$  is lost
since there are always two vortices in the trap,
and the forces exerted on the trap are always in the negative $x$ direction.
Figure~\ref{fig:12}(c) shows a transient situation at
$v_{tr} = 0.048$, where the trap is initially dragging
one vortex but then captures
a second vortex, as indicated by the drop in $f_x$ to a more
negative value.

\begin{figure}
\includegraphics[width=3.5in]{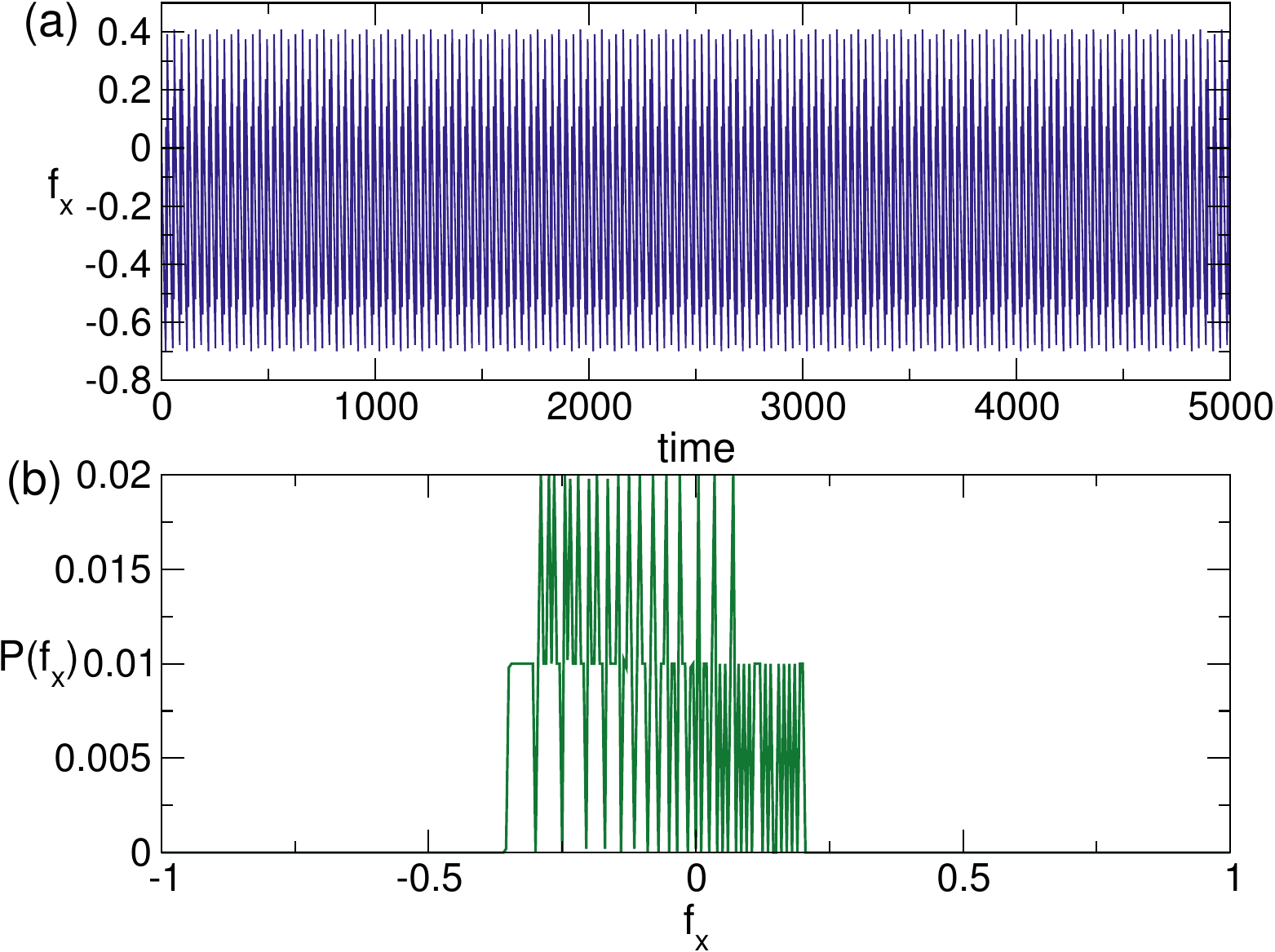}
\caption{ (a)
  $f_x(t)$ in phase I for a sample with
  $\theta = 0^\circ$,
  $F_{tr} = 1.0$, and $v_{tr} = 0.3$.
  (b) The corresponding $P(f_{x})$
  indicates that the
fluctuations are periodic.
}
\label{fig:13}
\end{figure}

In general, we find that when the trap is dragged
along certain symmetry angles of the pinning array,
such as $\theta=0^{\circ}$ and $\theta=45^{\circ}$,
the force fluctuations contain a stronger periodic component,
while for
driving at incommensurate angles,
the force fluctuations are more disordered.
Previous work with particles driven over square pinning arrays
showed that directional locking
should occur at angles of
$\theta = \tan^{-1}(n/m)$, where $n$ and $m$ are integers \cite{49,50,51,52,53},
so for driving at these locking angles,
we expect the force fluctuations to
be more periodic.
In Fig.~\ref{fig:13} we show $f_x(t)$ and $P(f_x)$ in
phase I for a sample with $\theta = 0^{\circ}$
at $F_{tr} = 1.0$ and
$v_{tr} = 0.3$.
There is a strong periodic signal
and $f_x(t)$ is much more ordered than the stick-slip time series shown
in Fig.~\ref{fig:10}(a) for a
$\theta=30^\circ$
system in phase I
at $F_{tr}=1.0$ and $v_{tr}=0.5$.

\section{Discussion}
Within the particle description we use, the
number of vortices
and their shape is held fixed;
however, it is possible that a sufficiently strong trapping force could induce
vortex shape distortions that could change the dynamics.
Additionally, if the trap is strong enough, then in
the multiple vortex trapping phases IV and V,
the trapped vortices may merge and form multiquantum states.
Our results
apply to the limit
in which the trap
is weak enough that such distortions do not occur.
In experiments, it is likely that the tip speed will be in the limit
of low $v_{tr}$;
however, the phase diagrams of Figs.~\ref{fig:6} and
\ref{fig:7} indicate that
most of the phases can be accessed even at the
lowest trap velocities by varying the trap strength.
It is also possible to use a stationary trap of fixed strength and apply a current
so that all of the vortices flow past the trap in order to exert forces on it.
When the vortices are moving fast enough that the trap cannot capture
a vortex, the system will be in the decoupled state.
Our results should be general to other systems of particles interacting with
periodic trap arrays, such as colloidal particles in optical or gravitational lattices,
where the interactions between colloids can be of magnetic form
with a $1/r^3$ behavior or of screened Coulomb or Yukawa form.

\section{Summary}
We have numerically examined vortex manipulation in superconductors
with a periodic array of pinning sites by a local moving trap.
We find five distinct phases
depending on the trap strength and velocity.
In phase I, which appears for
low trap strength
or large trap velocity, the vortices are decoupled from the trap, which can move
a vortex within a pinning site but cannot depin it.  The distribution of forces
experienced by the trap has a peak at zero force corresponding to time intervals
during which the trap is moving between adjacent pinning sites and contains
no vortex.
In the intermediate coupling phase II,
the trap drags a vortex
out of a pinning site
and then exchanges that vortex
with another vortex upon reaching the next occupied pinning site, so that
the trap is always occupied by a vortex.
In phase III, where intermediate trapping occurs,
the trap can drag a single vortex over
long distances, but still occasionally exchanges this vortex with another
pinned vortex.
Within phase III we find
a counterintuitive effect in which the trap couples more strongly to a single
vortex at higher velocities than at lower velocities, since at lower velocities
there is enough time for a pinned vortex to complete an exchange with the
trapped vortex.
Phases II and III both exhibit stick-slip fluctuations of the force experienced
by the trap
that correlate with vortex exchange events and with the entry and exit of
vortices from the trap.
Phase IV is an intermittent multiple trapping regime
in which the trap alternates between
capturing
one or two vortices,
producing a telegraph noise signature in the trap force fluctuation signal.
In phase V, where the trap is strongly coupled and always captures two vortices,
the telegraph noise signal is lost.
We map the evolutions of these phases for
varied trap coupling strength,
trap velocity, and the angle of trap motion with respect to
the $x$ symmetry axis of the pinning array.
For a given trap coupling force,
transitions among the phases
occur as a function of increasing trap velocity.
Our results should be general to other
types of particle systems with a periodic substrate
subjected to
a moving local trap, such
as colloidal particles, skyrmions, or ions on optical traps.

\acknowledgments
This work was carried out under the auspices of the
NNSA of the
U.S. DoE
at
LANL
under Contract No.
DE-AC52-06NA25396.


\begin{thebibliography}{99}
\bibitem{1}
  M. Baert, V.V. Metlushko, R. Jonckheere, V.V. Moshchalkov, and Y. Bruynseraede,
  Composite flux-line lattices stabilized in superconducting films by a regular
  array of artificial defects, Phys. Rev. Lett. {\bf 74}, 3269 (1995).

\bibitem{2}
  C. Reichhardt, C.J. Olson, and F. Nori,
  Commensurate and incommensurate vortex states in superconductors
  with periodic pinning arrays, Phys. Rev. B {\bf 57}, 7937 (1998).

\bibitem{3}
  G.R. Berdiyorov, M.V. Milosevic, and F.M. Peeters,
  Novel commensurability effects in superconducting films with antidot arrays,
  Phys. Rev. Lett. {\bf 96}, 207001 (2006).

\bibitem{4}
  M.V{\' e}lez, J.I.Mart{\' \i}n, J.E.Villegas, A.Hoffmann, E.M.Gonz{\' a}lez, J.L.Vicent,
  and I.K.Schuller,
Superconducting vortex pinning with artificial magnetic nanostructures,
J. Magn. Magn. Mater. {\bf 320}, 2547 (2008).

\bibitem{5}
  I.A. Sadovskyy, Y.L. Wang, Z.-L. Xiao, W.-K. Kwok, and A. Glatz,
  Effect of hexagonal patterned arrays and defect geometry on the critical
  current of superconducting films, Phys. Rev. B {\bf 95}, 075303 (2017).

\bibitem{6}
  D. S. Fisher, Collective transport in random media: From superconductors to
  earthquakes, Phys. Rep. {\bf 301}, 113 (1998).

\bibitem{7}
J. Gutierrez, A.V. Silhanek, J. Van de Vondel, W. Gillijns, and V.V. Moshchalkov,
  Transition from turbulent to nearly laminar vortex flow in superconductors with
  periodic pinning,
Phys. Rev. B {\bf 80}, 140514(R) (2009).

\bibitem{8}
  S. Avci, Z.L. Xiao, J. Hua, A. Imre, R. Divan, J. Pearson, U. Welp, W.K. Kwok,
  and G.W. Crabtree,
Matching effect and dynamic phases of vortex matter in Bi$_2$Sr$_2$CaCu$_2$O$_8$
nanoribbon with a periodic array of holes,
Appl. Phys. Lett. {\bf 97}, 042511 (2010).

\bibitem{9}
N. Poccia, T.I. Baturina, F. Coneri, C.G. Molenaar, X.R. Wang, G. Bianconi,
A. Brinkman, H. Hilgenkamp, A.A. Golubov, and V.M. Vinokur,
Critical behavior at a dynamic vortex insulator-to-metal transition,
Science {\bf 349}, 1202 (2015).

\bibitem{10}
  C. Reichhardt and C. J. Olson Reichhardt,
  Depinning and nonequilibrium dynamic phases of particle assemblies driven over random and ordered substrates: A review, Rep. Prog. Phys. {\bf 80}, 026501 (2017).

\bibitem{11}
  G. Blatter, M. V. Feigel'man, V.B. Geshkenbein, A.I. Larkin, and V.M. Vinokur,
  Vortices in high-temperature superconductors. Rev. Mod. Phys. {\bf 66},
  1125 (1994).

\bibitem{12}
  T. Giamarchi and P. Le Doussal, Phase diagrams of flux lattices with disorder,
  Phys. Rev. B {\bf 55}, 6577 (1997).

\bibitem{13}
  I. Guillam{\' o}n, R. C{\' o}rdoba, J. Ses{' e}, J.M. De Teresa, M.R. Ibarra, S. Viera,
  and H. Suderow,
  Enhancement of long-range correlations in a 2D vortex lattice by an incommensurate 1D disorder potential, Nature Phys. {\bf 10}, 851 (2014).

\bibitem{14}
C.J.O. Reichhardt and M.B. Hastings,
Do Vortices Entangle?
Phys. Rev. Lett. {\bf 92}, 157002 (2004).

\bibitem{15}
Y. Kafri, D. Nelson, and A. Polkovnikov,
Unzipping vortices in type-II superconductors,
Phys. Rev. B {\bf 76}, 144501 (2007).

\bibitem{16}
M.B. Hastings, C.J.O. Reichhardt, and C. Reichhardt,
Ratchet cellular automata,
Phys. Rev. Lett. {\bf 90}, 247004 (2003).

\bibitem{17}
M.V. Milosevic, G.R. Berdiyorov, and F.M. Peeters,
Fluxonic cellular automata,
Appl. Phys. Lett. {\bf 91}, 212501 (2007).

\bibitem{N}
D.A. Ivanov,
Non-Abelian statistics of half-quantum vortices in p-wave superconductors,
Phys. Rev. Lett. {\bf 86}, 268 (2001).

\bibitem{N2}
  H.-H. Sun {\it et al.},
  %K.-W. Zhang, L.-H. Hu, C. Li, G.-Y. Wang, H.-Y. Ma, Z.-A. Xu, C.-L. Gao, D.-D. Guan,
  %Y.-Y. Li, C. Liu, D. Qian, Y. Zhou, L. Fu, S.-C. Li, F.-C. Zhang, and J.-F. Jia,
  Majorana zero mode detected with spin selective Andreev reflection in the
  vortex of a topological superconductor,
Phys. Rev. Lett. {\bf 116}, 257003 (2016).

\bibitem{N.2}
H.-H. Sun and J.-F. Jia,
Detection of Majorana zero mode in the vortex,
npj Quantum Materials {\bf 2}, 34 (2017).

\bibitem{18}
Q.-F. Liang, Z. Wang, and X. Hu,
Manipulation of Majorana fermions by point-like gate voltage in the
vortex state of a topological superconductor,
EPL {\bf 99}, 50004 (2012).

\bibitem{New1}
    H.-D. Wu and T. Zhou,
    Vortex pinning by the point potential in topological superconductors: a
    scheme for braiding Majorana bound states,
arXiv:1710.04421.

\bibitem{19}
  B.W. Gardner, J.C. Wynn, D.A. Bonn, R. Liang, W.N. Hardy, J.R. Kirtley, V.G. Kogan,
  and K.A. Moler,
  Manipulation of single vortices in YBa$_2$Cu$_3$O$_{6.354}$ with a locally applied magnetic field,
  Appl. Phys. Lett. {\bf 80}, 1010 (2002).

\bibitem{20}
E.W.J. Straver, J.E. Hoffman, O.M. Auslaender, D. Rugar, and K.A. Moler,
Controlled manipulation of individual vortices in a superconductor,
Appl. Phys. Lett. {\bf 93}, 172514 (2008).

\bibitem{21}
C. Reichhardt,
High-temperature superconductors: Vortices wiggled and dragged,
Nature Phys. {\bf 5}, 15 (2009).

\bibitem{22}
  O.M. Auslaender, L. Luan, E.W.J. Straver, J.E. Hoffman, N.C. Koshnick, E. Zeldov,
  D.A. Bonn, R. Liang, W.N. Hardy, and K.A. Moler,
Mechanics of individual isolated vortices in a cuprate superconductor,
Nature Phys. {\bf 5}, 35 (2009).

\bibitem{23}
L. Luan, O.M. Auslaender, D.A. Bonn, R. Liang, W.N. Hardy, and K.A. Moler,
Magnetic force microscopy study of interlayer kinks in individual vortices in the
underdoped cuprate superconductor YBa$_2$Cu$_3$O$_{6+x}$,
Phys. Rev. B {\bf 79}, 214530 (2009).

\bibitem{24}
  N. Shapira, Y. Lamhot, O. Shpielberg, Y. Kafri, B. J. Ramshaw, D.A. Bonn, R. Liang,
  W.N. Hardy, and O.M. Auslaender,
Disorder-induced power-law response of a superconducting vortex on a plane,
Phys. Rev. B {\bf 92}, 100501(R) (2015).

\bibitem{25}
  I.S. Veshchunov, W. Magrini, S.V. Mironov, A.G. Godin, J.-B. Trebbia, A.I. Buzdin,
  Ph. Tamarat, and B. Lounis,
Optical manipulation of single flux quanta,
Nature Commun. {\bf 7}, 12801 (2016).

\bibitem{26}
A. Kremen, S. Wissberg, N. Haham, E. Persky, Y. Frenkel, and B. Kalisky,
Mechanical control of individual superconducting vortices,
Nano Lett. {\bf 16}, 1626 (2016).

\bibitem{27}
  J.-Y. Ge, V.N. Gladilin, J. Tempere, C. Xue, J.T. Devreese, J. Van de Vondel,
  Y. Zhou, and V.V. Moshchalkov,
Nanoscale assembly of superconducting vortices with scanning tunnelling microscope tip,
Nature Commun. {\bf 7}, 13880 (2016).

\bibitem{28}
J.-Y. Ge, V.N. Gladilin, J. Tempere, J. Devreese, and V.V. Moshchalkov,
Controlled generation of quantized vortex-antivortex pairs in a superconducting condensate,
Nano Lett. {\bf 17}, 5003 (2017).

\bibitem{29}
M.B. Hastings, C.J.O. Reichhardt, and C. Reichhardt,
Depinning by fracture in a glassy background,
Phys. Rev. Lett. {\bf 90}, 098302 (2003).

\bibitem{30}
P. Habdas, D. Schaar, A.C. Levitt, and E.R. Weeks,
Forced motion of a probe particle near the colloidal glass transition,
Europhys. Lett. {\bf 67}, 477 (2004).

\bibitem{31}
C. Reichhardt and C.J.O. Reichhardt,
Crossover from intermittent to continuum dynamics for locally driven colloids,
Phys. Rev. Lett. {\bf 96}, 028301 (2006).

\bibitem{32}
C.J.O. Reichhardt and C. Reichhardt,
Viscous decoupling transitions for individually dragged particles in systems with quenched disorder,
Phys. Rev. E {\bf 78}, 011402 (2008).

\bibitem{33}
D. Winter, J. Horbach, P. Virnau, and K. Binder,
Active nonlinear microrheology in a glass-forming Yukawa fluid,
Phys. Rev. Lett. {\bf 108}, 028303 (2012).

\bibitem{34}
Th. Voigtmann and M. Fuchs,
Force-driven micro-rheology,
Eur. Phys. J. Spec. Top. {\bf 222}, 2819 (2013).

\bibitem{35}
C. Reichhardt and C.J.O. Reichhardt,
Local melting and drag for a particle driven through a colloidal crystal,
Phys. Rev. Lett. {\bf 92}, 108301 (2004).

\bibitem{36}
R.P.A. Dullens and C. Bechinger,
Shear thinning and local melting of colloidal crystals,
Phys. Rev. Lett. {\bf 107}, 138301 (2011).

\bibitem{37}
  K. Harada, O. Kamimura, H. Kasai, T. Matsuda, A. Tonomura, and V.V. Moshchalkov,
  Direct observation of vortex dynamics in superconducting films with regular arrays of defects,
  Science {\bf 274}, 1167 (1996).

\bibitem{38}
  S. Tung, V. Schweikhard, and E.A. Cornell,
  Observation of vortex pinning in Bose-Einstein condensates,
  Phys. Rev. Lett. {\bf 97}, 240402 (2006).

\bibitem{39}
C. Reichhardt, D. Ray, and C.J.O. Reichhardt,
Quantized transport for a skyrmion moving on a two-dimensional periodic substrate,
Phys. Rev. B {\bf 91}, 104426 (2015).

\bibitem{40}
  A. Benassi, A. Vanossi, and E. Tosatti, Nanofriction in cold ion traps, Nature Commun.
  {\bf 2}, 236 (2011).

\bibitem{41}
P.T. Korda, M.B. Taylor, and D.G. Grier,
Kinetically locked-In colloidal transport in an array of optical tweezers,
Phys. Rev. Lett. {\bf 89}, 128301 (2002).

\bibitem{42}
T. Bohlein, J. Mikhael, and C. Bechinger,
Observation of kinks and antikinks in colloidal monolayers driven across ordered surfaces,
Nature Mater. {\bf 11}, 126 (2012).

\bibitem{43}
A. Vanossi, N. Manini, and E. Tosatti,
Static and dynamic friction in sliding colloidal monolayers,
Proc. Natl. Acad. Sci. (U.S.A.) {\bf 109}, 16429 (2012).

\bibitem{44}
D. McDermott, J. Amelang, C. J. Olson Reichhardt, and C. Reichhardt,
Dynamic regimes for driven colloidal particles on a periodic substrate at commensurate and incommensurate fillings,
Phys. Rev. E {\bf 88}, 062301 (2013).

\bibitem{45}
  A. Vanossi, N. Manini, M. Urbakh, S. Zapperi, and E. Tosatti,
  Modeling friction: From nanoscale to mesoscale, Rev. Mod. Phys. {\bf 85}, 529 (2013).

\bibitem{46}
C. Reichhardt, C. J. Olson, and F. Nori,
Nonequilibrium dynamic phases and plastic flow of driven vortex lattices in
superconductors with periodic arrays of pinning sites,
Phys. Rev. B {\bf 58}, 6534 (1998).

\bibitem{47}
V. Misko, S. Savel’ev, A. Rakhmanov, and F. Nori,
Negative differential resistivity in superconductors with periodic arrays of pinning sites,
Phys. Rev. B {\bf 75}, 024509 (2007).

\bibitem{48}
C. Reichhardt and C.J.O. Reichhardt,
Moving vortex phases, dynamical symmetry breaking, and jamming for vortices in
honeycomb pinning arrays,
Phys. Rev. B {\bf 78}, 224511 (2008).

\bibitem{49}
C. Reichhardt and F. Nori,
Phase locking, devil's staircases, Farey trees, and Arnold tongues in
driven vortex lattices with periodic pinning,
Phys. Rev. Lett. {\bf 82}, 414 (1999).

\bibitem{50}
J. M. Sancho, M. Khoury, K. Lindenberg, and A. M. Lacasta,
Particle separation by external fields on periodic surfaces,
J. Phys.: Condens. Matter {\bf 17}, S4151 (2005).

\bibitem{51}
J. Villegas, E. Gonzalez, M. Montero, I. Schuller, and J. Vicent,
Vortex-lattice dynamics with channeled pinning potential landscapes,
Phys. Rev. B {\bf 72}, 064507 (2005).

\bibitem{52}
J. Koplik and G. Drazer,
Nanoscale simulations of directional locking,
Phys. Fluids {\bf 22}, 052005 (2010).

\bibitem{53}
C. Reichhardt and C. J. O. Reichhardt,
Structural transitions and dynamical regimes for directional locking of vortices and
colloids driven over periodic substrates,
J. Phys.: Condens. Matter {\bf 24}, 225702 (2012).

\end{thebibliography}
\end{document}